\documentclass [aps,12pt,aps,a4paper,prb]{revtex4}
\input epsf
\usepackage{graphicx}
\usepackage{amssymb}
\usepackage{latexsym}

\begin{document}
%\renewcommand{\topfraction}{0.9}
%\renewcommand{\textfraction}{0.1}
%\renewcommand{\floatpagefraction}{0.75}
%\topmargin 2.0
%\newcommand {\beq} { \begin{equation} }
%\newcommand {\eeq} { \end{equation} }
%\newcommand {\dis} { \displaystyle }
%\newcommand {\fle} { \stackrel{\rightarrow} }
%\newcommand {\som} { \stackrel{\wedge} }
%\newcommand {\bea} { \begin{array} }
%\newcommand {\ear} { \end{array} }
%\newcommand {\iz} {\left}
%\newcommand {\de} {\right}
\newcommand {\etal} { \emph{et al }}
\section*{}
\begin{center}
\title{\Large\bf Microphase Separation in Polyelectrolytic Diblock Copolymer Melt : Weak Segregation Limit}
\author{{\bf Rajeev Kumar and M.Muthukumar} \footnote[1]{To whom any correspondence should be
addressed, Email : muthu@polysci.umass.edu}}

\affiliation{\it Dept. of Polymer Science \& Engineering, Materials Research Science \& Engineering Center,\\
 University of Massachusetts,
 Amherst, MA-01003, USA.}
\maketitle

\end{center}
\vskip0.1cm
\begin{center}
ABSTRACT
\end{center}
\vskip0.1cm

\noindent We present a generalized theory of microphase separation
for charged-neutral diblock copolymer melt. Stability limit of the disordered phase for salt-free melt has been calculated using Random Phase Approximation (RPA) and self-consistent field theory (SCFT). Explicit analytical free energy expressions for different classical ordered microstructures (lamellar, cylinder and sphere) are presented. We demonstrate that chemical mismatch required for the onset of microphase separation ($\chi^{\star} N$) in charged-neutral diblock melt is higher and the period of ordered microstructures is lower than those for the corresponding neutral-neutral diblock system. Theoretical predictions on the period of ordered structures in terms of Coulomb electrostatic interaction strength, chain length, block length, and the chemical mismatch between blocks are presented. SCFT has been used to go beyond the stability limit, where electrostatic potential and charge distribution are calculated self-consistently. Stability limits calculated using RPA are in perfect agreement with the corresponding SCFT calculations. Limiting laws for stability limit and the period of ordered structures are presented and comparisons are made with an earlier theory. Also, transition boundaries between different morphologies have been investigated. 
\clearpage
 \section{INTRODUCTION}
\setcounter {equation} {0}

Science behind the complex behavior of amphiphilic systems continues to be of interest to scientific community. A great deal
of theoretical\cite{khokhlov80,borue88,marko91,vilgis92,bouayed92,
khokhlov94,muthu99, muthu02} and experimental
efforts\cite{kanaya011,kanaya01,eisenberg98,jenekhe98,jenekhe99,kataoka01} have been made to study
this behavior. Especially, self-assembly of
amphiphiles\cite{eisenberg98} is of significant importance in
understanding many biological systems. An amphiphilic diblock copolymer system has applications
such as encapsulation\cite{jenekhe98,jenekhe99} and drug
delivery\cite{kataoka01}, which are dependent on self-assembly of macromolecules. 

Self-assembly of neutral block copolymers in concentrated regimes, 
has already been studied
extensively in the last three decades. Seminal work in developing theory
for microdomains in diblock copolymer melt was done by Helfand\cite{helfand71} {\etal},
where unit cell approximation was used to calculate the properties for sharp interfaces 
(strong segregation limit (SSL)). Later on, Leibler \cite{leibler80}
calculated morphology diagrams in weak segregation limit (WSL) 
using the random phase approximation (RPA), when the system is on the verge 
of transformation from disordered to ordered phases 
and interfaces between the ordered domains are diffuse.
Few years later, Semenov \cite{semenov84} and Ohta
\cite{ohta86}{\etal} calculated morphology
diagrams in SSL taking into account a sharp interface between the domains. Semenov used ground state dominance and Ohta {\etal} extended the RPA method developed by Leibler to SSL. Afterwards, Muthukumar
\cite{dft91,dft93} {\etal} and Matsen\cite{matsen94}{\etal} bridged the gap between WSL and SSL theories by using 
density functional theory (DFT) and SCFT respectively. To go beyond mean field, fluctuations of order 
parameter were included by Fredrickson\cite{helfand87} {\etal}, Olvera de la cruz\cite{cruz91} and Muthukumar\cite{muthu93}.

Although we have a sound understanding of neutral
copolymers, our understanding of charged copolymers is
inadequate.  A number of researchers have tried to explore charged
diblock systems. Some  of these efforts are invested in studying dilute
solutions (micelle regime)\cite{zhulina02,semenov05} and
others have been carried out in the concentrated
regime\cite{marko91,vilgis92,bouayed92,fraaije041,fraaije042}. The fundamental question is
how Coulomb interactions affect the relative stabilities of ordered morphologies. First work in this direction was carried out by Marko and Rabin \cite{marko91} 
who explored charged copolymers in both the melts and solutions. 
In their study, they presented the effect of degree of ionization and salt 
on critical parameters for weak segregation in melt and studied micelle behaviour also.
But they did not consider any ordered microstructures for melts. Recently, SCFT for polyelectrolytic systems
has been developed by Shi\cite{shi99}{\etal} and Wang\cite{fred04} {\etal}.

In this paper, we have considered microphase
separation in diblock copolymer melts, formed out of polyelectrolyte ($A$) block and neutral ($B$) block. RPA is
presented in Sec. ~\ref{sec:RPA} and SCFT equations are briefly presented in Sec. ~\ref{sec:SCFT}. Calculated results and 
conclusions are presented in Sec. ~\ref{sec:results} and ~\ref{sec:discuss}, respectively. 

\section{THEORY}
\setcounter {equation} {0} \label{sec:theory}

We consider a system of $n$ block copolymer chains, each
containing a total of $N$  segments of two species ($A$ and $B$) with
$f$ as the fraction of the block $A$. Number of segments in blocks $A$
and $B$ is represented by $N_{A}$ and  $N_{B}$, respectively, so that ($f = N_{A}/N, N = N_{A} + N_{B}$).
 Block $A$ is taken to be
polyelectrolytic (negatively charged) and we assume that there are $n_{c}$
counterions released by the charged block. In addition to this, there are $n_{\gamma}$ ions 
of species $\gamma$ coming from the added salt (in 
total volume $\Omega$ ) so that the whole system is globally electroneutral. Let $Z_{i}$ be 
the valency of the $i^{th}$ charged species. Subscripts $A,B,c,+$ and $-$ are used to represent 
monomer $A$, $B$, counterion from block $A$, positive and negative salt ions, respectively.
The degree of
ionization of the $A$ block per chain is taken to be $\alpha$ so that each of the $f \alpha N$ segments 
of $A$ block chain carry a charge of $eZ_{A}$, where $e$ is the electronic charge.
So, there are $N(1-f\alpha)$ uncharged segments in a chain. Recently, it has been shown that
specific charge distributions along the backbone play a significant role in the physics of polyelectrolytes\cite{fred04}. 
In our current study, we consider smeared charge distribution i.e. {\it each} segment of block $A$ has charge $e \alpha Z_{A}$.
We represent copolymer chain as a continuous curve
of length $Nl$, where $l$ is the Kuhn segment length. For
the treatment shown below, we assume that volume occupied by each $A$ and $B$ 
monomer is the same ($ = l^{3} \equiv 1/\rho_{o}, \rho_{o}$ being the bulk density) 
and that the system is incompressible so that $\Omega = n N l^{3}$. 
We use an arc length variable $s_{j}$
to represent any segment along the backbone of $j^{th}$ chain. In this notation, $A$ 
block in $j^{th}$ chain is represented by $s_{j} \: (0\leq s_{j}\leq N_{A}l)$. Also, 
the position vector for a particular segment is represented by $R_{j}(s_{j})$. 
For this system, Helmholtz free energy $F$ can be written within Edward's formalism by:
\begin{eqnarray}
       \mbox{exp}\left(-\frac{F} {k_{B}T}\right )& = & \frac {1}{n!n_{c}!\prod_{\gamma}
       n_{\gamma}!}\int\prod_{j=1}^{n} D[R_{j}]\int
       \prod_{i=1}^{n_{c}+ \sum_{\gamma}n_{\gamma}} dr_{i}
       \quad \mbox{exp} \left \{-\frac {3}{2 l}\sum_{j = 1}^{n}\int_{0}^{Nl}
       {ds_{j}}\left(\frac{\partial R_{j}}
       {\partial s_{j}} \right )^{2} \right . \nonumber \\
       & & -  \frac {1}{2 {l}^{2}} \sum_{j =1}^{n}\sum_{k =1}^{n}
       \left [\int_{0}^{N_{A}l}{ds_{j}} \int_{0}^{N_{A}l}{ds_{k}}
       V_{AA} [ R_{j}(s_{j}) - R_{k}(s_{k})] \right . \nonumber \\
       & & + \int_{N_{A}l}^{Nl}{ds_{j}} \int_{N_{A}l}^{Nl}{ds_{k}}
       V_{BB} [ R_{j}(s_{j}) - R_{k}(s_{k})]\nonumber \\
       && \left . + 2 \int_{0}^{N_{A}l}{ds_{j}}\int_{N_{A}l}^{Nl}{ds_{k}}
       V_{AB} [ R_{j}(s_{j})- R_{k}(s_{k})] \right ] \nonumber \\
       & & -\frac{1}{{l}}\sum_{j =1 }^{n}\sum_{m =1 }^{n_{c}+ \sum_{\gamma}n_{\gamma}}\left [
       \int_{0}^{N_{A}l}{ds_{j} }V_{Am} [ R_{j}(s_{j}) - r_{m}]
       + \int_{N_{A}l}^{Nl}{ds_{j} }V_{Bm} [ R_{j}(s_{j}) - r_{m}] \right] \nonumber \\
       & & - \left . \frac{1}{2}\sum_{m =1 }^{n_{c}+ \sum_{\gamma}n_{\gamma}}\sum_{p =1 }^{n_{c}+ \sum_{\gamma}n_{\gamma}}
       V_{mp} [ r_{m} - r_{p}] \right \}\prod_{r}\delta\left[ \hat{\rho}_{A}(r) + \hat{\rho}_{B}(r) -\rho_{0} \right ]
       \label{eq:parti}
\end{eqnarray}
Here, $k_{B}T$ is Boltzmann constant times the absolute
temperature. $V_{AA}(r),V_{BB}(r) \mbox{ and}\: V_{AB}(r)$ are the
interaction energies between segments of different types separated
by distance $\mathbf{r}= \mid r \mid $ so that
\begin{eqnarray}
       V_{AA}(r) & = & w_{AA}\delta(\mathbf{r}) + \frac {Z_{A}^{2}e^{2}\alpha^{2}}{\epsilon
       k_{B}T}\frac{1}{\mathbf{r}} \\
       V_{BB}(r) & = & w_{BB}\delta(\mathbf{r}) \\
       V_{AB}(r) & = & w_{AB}\delta(\mathbf{r})
\end{eqnarray}
where $w_{AA}$, $w_{BB}$ and $w_{AB}$ are the intersegment excluded
volumes arising from the short range interactions,
$\delta(\mathbf{r})$ is Dirac delta function and $\epsilon$ is
{\it position independent effective} dielectric constant 
of the medium ( in units of $ 4 \pi \epsilon_{o}$ where $\epsilon_{o}$ is 
the permittivity of vacuum). In writing $V_{AA}(r)$, it is assumed that 
total charge of any chain is uniformly distributed along $A$
block (smeared charge distribution). 
Interactions between polymer segments and small charged
molecules (counterions and coions) are accounted for by
treating small ions as point charges so that they have zero excluded
volume and are purely electrostatic in nature. These
interactions, represented by $V_{Am}(r)$, $V_{Bm}(r)$ and
$V_{mp}(r)$ in Eq. (~\ref{eq:parti}), are given by
\begin{eqnarray}
       V_{Am}(r) & = & \frac {Z_{A} Z_{m} e^{2} \alpha }{\epsilon
       k_{B}T}\frac{1}{\mathbf{r}} \\
       V_{Bm}(r) & = & 0 \\
       V_{mp}(r) & = & \frac {Z_{m}Z_{p} e^{2}}{\epsilon k_{B}T}\frac{1}{\mathbf{r}}
\end{eqnarray}
The Dirac delta function involving $\hat{\rho}_{A}(r) $ and $\hat{\rho}_{B}(r)$ enforce the incompressibility of the system at all locations, where $\hat{\rho}_{A}(r) $ and $\hat{\rho}_{B}(r)$ are {\it microscopic} monomer number densities defined as
 \begin{eqnarray}
    \hat{\rho}_{A}(r) & = & \frac{1}{l}\sum_{j=1}^{n}\int_{0}^{N_{A}l}{ds_{j}}\, \delta (r-R_{j}(s_{j})) \label{eq:density} \\
    \hat{\rho}_{B}(r) & = & \frac{1}{l}\sum_{j=1}^{n}\int_{
    N_{A}l}^{Nl}{ds_{j}}\, \delta (r-R_{j}(s_{j}))
 \end {eqnarray}

\subsection{WSL - Random Phase Approximation} \label{sec:RPA}
\subsubsection{Stability Limit }

Near the stability limit of the melt, densities of all components in the inhomogeneous phase deviate slightly from their average values in the homogeneous phase. So, the free energy of the inhomogenous phase can be obtained by expanding the corresponding free energy expression for the homogeneous phase about the average densities. Neglecting cubic and higher order terms in densities, the free energy of the inhomogeneous phase is obtained in terms of the order parameter $\phi(r)$ (see Appendix A) and  has the form:
\begin{equation}
F =  F_{0} + \delta F(\phi) + O(\phi^{3}) \label{eq:flory},
\end{equation}
where the order parameter $\phi$ is the same as that used for neutral block copolymer melts, given by
\begin{equation}
      \phi(r)= \, <\delta \hat{\rho}_{A}(r)> \, = \, <\hat{\rho}_{A}(r)> - \, f\,\rho_{0}.
 \end{equation}
$F_{0}$ is the free energy of the homogeneous phase and is
given by a Flory-type equation:
\begin{equation}
\frac{F_{0}}{k_{B}T} =  \sum_{i = c,+,-} n_{i}\ln\left
(n_{i}\right) - \frac{\kappa^{3}}{12\pi} + \chi_{AB}f(1-f) \label{eq:homo}
\end{equation}
In writing $F_{0}$, linear and constant terms have been ignored. In Eq. (~\ref{eq:homo}), $\chi_{AB}$ and $\kappa^{-1} = r_{D}$ are Flory's chi parameter and the Debye screening length, respectively. $r_{D}$ is given by 
 \begin{eqnarray}
    \kappa ^{2} & = & 4\pi l_{B}\left (\frac {n_{c}}{\Omega}Z_{c}^{2}+ \sum_{\gamma = +,-}\frac {n_{\gamma}}{\Omega}Z_{\gamma}^{2} \right )
 \end {eqnarray}
and $l_{B}= e^{2}/\epsilon k_{B}T $ stands for the Bjerrum length (in units of $4 \pi \epsilon_{o}$). In Eq. (~\ref{eq:homo}), the first term corresponds to the translational 
entropy of small ions, $\kappa^{3}$ term
accounts for counterions and coions correlation effect and $\chi_{AB}$ term gives the contribution of chemical mismatch between 
blocks.  

The second degree term in $\phi$ defines the structure factor ($S(k)$) and is given by:
\begin{eqnarray}
\delta F(\phi)&= & \frac{1}{2}\int \frac{d^{3}k}{(2\pi)^{3}} S^{-1}(k)\phi(k)\phi(-k) \label{eq:deltaf}\\
S^{-1}(k)& = & S_{0}^{-1}(k)+  S_{1}^{-1}(k) \label{eq:struct}\\
S_{0}^{-1}(k)&=& Q(x) - 2\chi_{AB}l^{3} \\
Q(x) &=& \frac{g(1,x)}{\rho_{0}N\left\{g(f,x)g(1-f,x)-[g(1,x)-g(f,x)-g(1-f,x)]^{2}/4\right \}} \label{eq:q_func}\\
S_{1}^{-1}(k)&= & \frac{4\pi l_{B} Z_{A}^{2}\alpha^{2}N l^{2}}{6 x + \kappa^{2}N l^{2}} \label{eq:chargestruct}
\end {eqnarray}
In Eq. (~\ref{eq:struct}), $S_{0}^{-1}(k)$ is the contribution due to short range excluded volume interactions 
and $S_{1}^{-1}(k)$ is due to the long range Coulomb interactions present in
the system. $g(f,x)$ is the Debye function given by 
\begin{equation}
g(f,x)= \frac{2(fx + e^{-fx}-1)}{x^{2}},\quad x = \frac{k^{2}N l^{2}}{6}= k^{2}R_{g}^{2}  
\label{eq:debye}
\end{equation}
Note that Eqs. (~\ref{eq:deltaf} - ~\ref{eq:chargestruct}) are the same as in Ref.\cite{marko91}.

\subsubsection{Limiting Laws}
 To understand the qualitative behaviour of the system at the stability limit, a scaling analysis is presented here for two limiting cases. In order to carry out the scaling analysis, the complicated equation for  $S_{0}^{-1}(k)$ is  approximated by an expression,
 used previously for neutral block copolymers\cite{ohta86}, which reproduces $S_{0}^{-1}(k)$
 with about 5 \% accuracy. Using the approximate expression, Eq. (~\ref{eq:deltaf}) becomes:
\begin{eqnarray}
 \delta F\{\phi\}&=& \frac {1}{2\rho_{0}N}\int
\frac{d^{3}k}{(2\pi)^{3}}\left \{B(f)k^{2}+ \frac{A(f)}{k^{2}}+
\frac{C(l_{B})}{k^{2}+
\kappa^{2}}- \bar{\chi}\right \}\phi(k)\phi(-k)\\
B(f)\quad &= & \frac{N l^{2}}{12 f (1-f)}\\
A(f)\quad &= & \frac{9}{N l^{2}f^{2}(1-f)^{2}}\\
C(l_{B})&=& 4\pi l_{B} Z_{A}^{2}\alpha^{2}\rho_{0}N \\
\bar{\chi}\qquad &= & 2 \chi_{AB}N\rho_{0}l^{3} - \frac
{s(f)}{2f^{2}(1-f)^{2}}
\end {eqnarray}
Here, $s(f)$ is a parameter used to reproduce $S_{0}^{-1}(k)$.
At the stability limit, $S(k)$ diverges at the wave-vector $k = k^{\star}$ and the second degree term $\delta F$ in free energy vanishes at $k = k^{\star}\,\mbox{and}\,\chi_{AB}N = \chi_{AB}^{\star}N$. In the case of neutral block copolymers (i.e. $C(l_{B})=0$), the divergence corresponds to $k^{\star} = (A/B)^{1/4} \sim N^{-1/2}$ so that the period $D = 2\pi/k^{\star} \sim N^{1/2}$. Hence, $\delta F$ vanishes at $\bar{\chi}^{\star} = 2\sqrt{AB}$ so that $\chi_{AB}^{\star}N$ is given by expression
\begin{equation}
\left (\chi_{AB}^{\star}N \right )_{neutral} =\frac{1}{\rho_{0}l^{3}} \left [\frac{s(f)}{4 f^{2}(1-f)^{2}} + \sqrt{\frac{3}{4f^{3}(1-f)^{3}}}\,\,  \right ] \label{eq:neutralchi}
\end{equation} 
Now, for charged block copolymers, consider the two limiting cases: (i) $k^{2}\ll \kappa^{2}$  and (ii) $k^{2}\gg \kappa^{2}$. In terms of the Debye screening length , these cases correspond to $r_{D}\ll D$ and $r_{D}\gg D$, respectively.
\begin{center} Case (i): \, $r_{D}\ll D$ \end{center}
In this regime, the structure factor becomes
\begin{equation}
S^{-1}(k) \simeq B k^{2} + \frac{A}{k^{2}} - \left( \bar{\chi} - \frac{C}{\kappa^{2}} \right )
\end{equation} 
The maxima of the structure factor corresponds to $k^{\star} = (A/B)^{1/4}\sim N^{-1/2}$ so that the period $D = 2\pi/k^{\star} \sim N^{1/2}$ and is {\it independent} of degree of ionization of charged block. So,charged block copolymer behaves like neutral copolymer. This is because the electrostatic interactions are short ranged in this regime. But there is a remarkable effect of small ions on $\chi_{AB}^{\star} N$ and in fact, it is found that charged block copolymers have to have higher $\chi_{AB}^{\star} N$ as compared to its neutral analog for undergoing the microphase separation. For salty systems,
\begin{equation}
\left (\chi^{\star}_{AB}N \right )_{charged} = \left (\chi^{\star}_{AB}N \right )_{neutral} + \frac{2 \pi l_{B} Z_{A}^{2}\alpha^{2}\rho_{0}N}{\kappa^{2}}
\end{equation}
 For salt-free incompressible system where $\kappa^{2} = 4\pi l_{B}f\alpha Z_{c}^{2}/l^{3}$, this expression simplifies to
\begin{equation}
\left (\chi^{\star}_{AB}N \right )_{charged} = \left (\chi^{\star}_{AB}N \right )_{neutral} + \frac{1}{2}\left (\frac{Z_{A}}{Z_{c}}\right )^{2} \frac{\alpha N }{f}
\end{equation}     
Physically, this means that homogeneous phase in charged copolymer enjoys larger parameter space as compared to its neutral analog.
\begin{center} Case (ii):\, $r_{D}\gg D$ \end{center}
In this regime, the structure factor becomes
\begin{equation}
S^{-1}(k) = B k^{2} + \frac{A + C}{k^{2}} - \bar{\chi} 
\end{equation} 
The maxima of the structure factor corresponds to 
\begin{equation}
k^{\star} = \left (\frac{A+C}{B}\right )^{1/4} \Rightarrow D = 2 \pi\left (\frac{A+C}{B}\right )^{-1/4}
\end{equation} 
so that $\bar{\chi}^{\star} = 2\sqrt{(A+C)B}$. Plugging in expressions for A, B and C
\begin{eqnarray}
D &=& \frac{2\pi N^{1/2}f^{1/4}(1-f)^{1/4}l}{\left [108 + 48\pi Z_{A}^{2}(1-f)^{2}(f \alpha N)^{2}l_{B}l^{2}\rho_{0}\right ]^{1/4} }\\
\left ( \chi_{AB}^{\star}N \right )_{charged} &=& \frac{1}{\rho_{0}l^{3}}\left [ \frac{s(f)}{4 f^{2}(1-f)^{2}} + \sqrt{\frac{3}{4f^{3}(1-f)^{3}}+ \frac{(4\pi l_{B} l^{2}\rho_{0})(\alpha N)^{2}Z_{A}^{2}}{12 f(1-f)}}\,\,  \right ]\label{eq:chargedchi}
\end{eqnarray}
%For salt-free systems, the above expression can be rewritten in terms of $\kappa$ as 
%\begin{eqnarray}
%D&=& \frac{2\pi N^{1/2}f^{1/4}(1-f)^{1/4}l}{108^{1/4}\left [ 1 + \frac{Z_{A}^{2}}{9 Z_{c}^{2}}f^{2}(1-f)^{2}(f\alpha N)(\kappa^{2}Nl^{2})(\rho_{0} l^{3})\right ]^{1/4} } \label{eq:dcharge} \\
%\left ( \chi_{AB}^{\star}N \right )_{charged} &=& \frac{s(f)}{4 f^{2}(1-f)^{2}} + \frac{1}{\rho_{0}l^{3}}\sqrt{\frac{3}{4f^{3}(1-f)^{3}}+ \frac{(4\pi l_{B} l^{2}\rho_{0})(f\alpha N)^{2}Z_{A}^{2}}{12 f(1-f)}}  \label{eq:chargedchi}
%\end{eqnarray}
From the expression for $D$, it is clear that {\it the period decreases with an increase in the degree of ionization } in this limiting case. Comparing $\chi_{AB}^{\star}N$ for charged and neutral block copolymer cases (Eq. (~\ref{eq:chargedchi}) and Eq. (~\ref{eq:neutralchi})), it can be inferred that in this limiting regime also, $\left (\chi_{AB}^{\star}N \right )_{charged}> \left (\chi_{AB}^{\star}N \right )_{neutral}$ and increases with an increase in $\alpha$.

Further, it can be shown that the correlation effect of ions (Debye-H\"{u}ckel theory) is weak as long as $\kappa l_{B}\ll 1$. This means that all of the above limiting laws for salt-free systems are valid as long as $\kappa l_{B}\ll 1$.

\subsubsection{Numerical Calculations for Stability Limit}
From here onwards, we consider the salt-free melt. To calculate the stability limit for salt-free melts, inverse temperature dependence of $\chi_{AB}$ and $l_{B}$ is clubbed together by the introduction of a parameter (called reduced temperature\cite{muthu021}) defined as
\begin{equation}
t= \frac{l}{4\pi \alpha l_{B} } \quad \mbox{and}\quad \chi_{AB} = \frac{1}{20\pi t} \label{eq:reduced}
\end{equation}

\begin{figure}[h]
   \begin{center}
    \vspace*{1.0cm}
   \begin{minipage}[c]{15cm}
   {\epsfxsize= 15cm \epsfbox{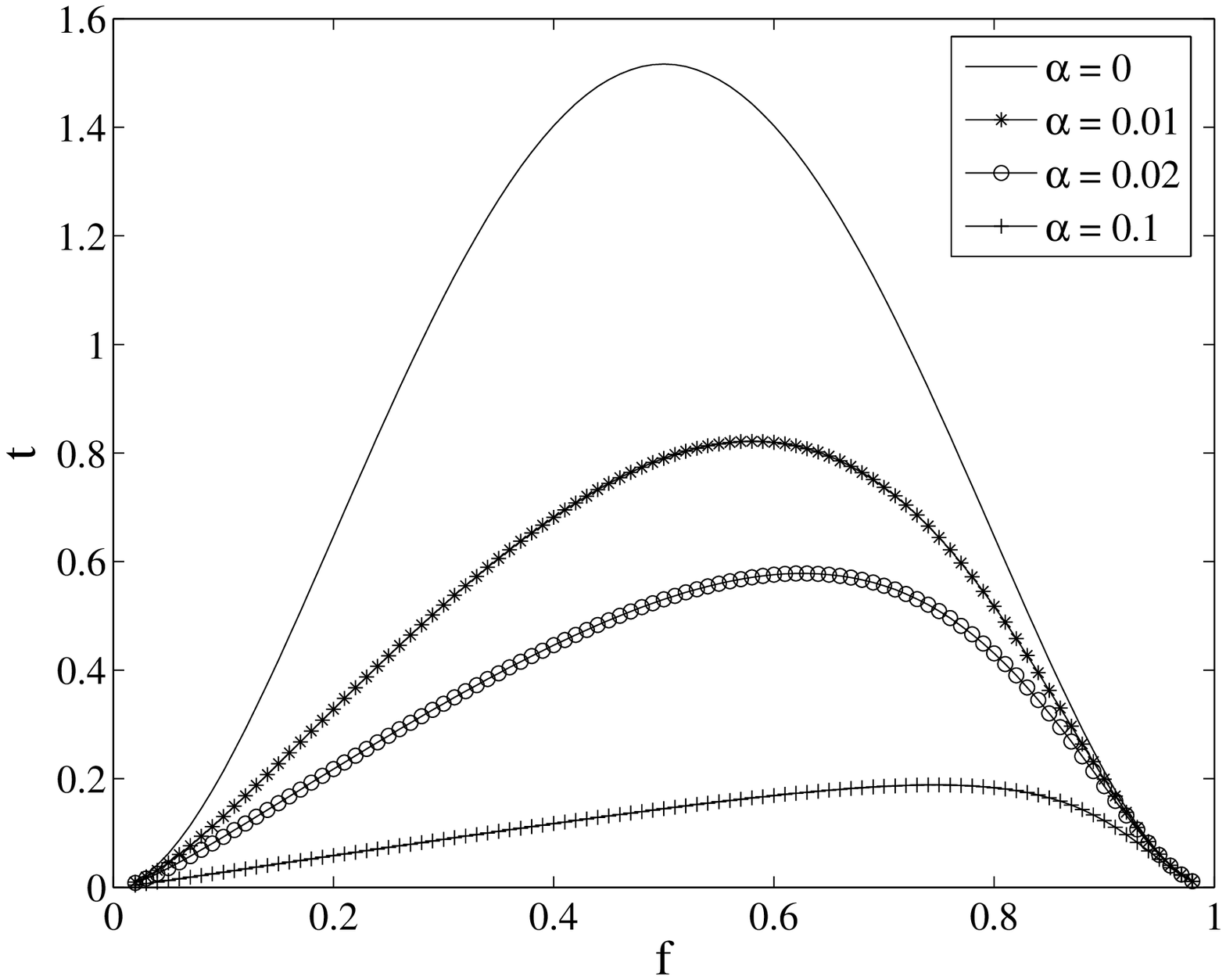}}
   %\label{fig:stability_n1000}
  \end{minipage}
\caption{} \label{stability_n1000}
%\caption{ Effect of degree of ionization ($\alpha$) on the stability limit in polyelectrolytic diblock
%melt: plots correspond to $N = 1000$ and $\alpha = 0, 0.01, 0.05, 0.1 $.}
 \end{center}
\end{figure}

\begin{figure}[h]
   \begin{center}
    \vspace*{1.0cm}
   \begin{minipage}[c]{15cm}
   {\epsfxsize= 15cm \epsfbox{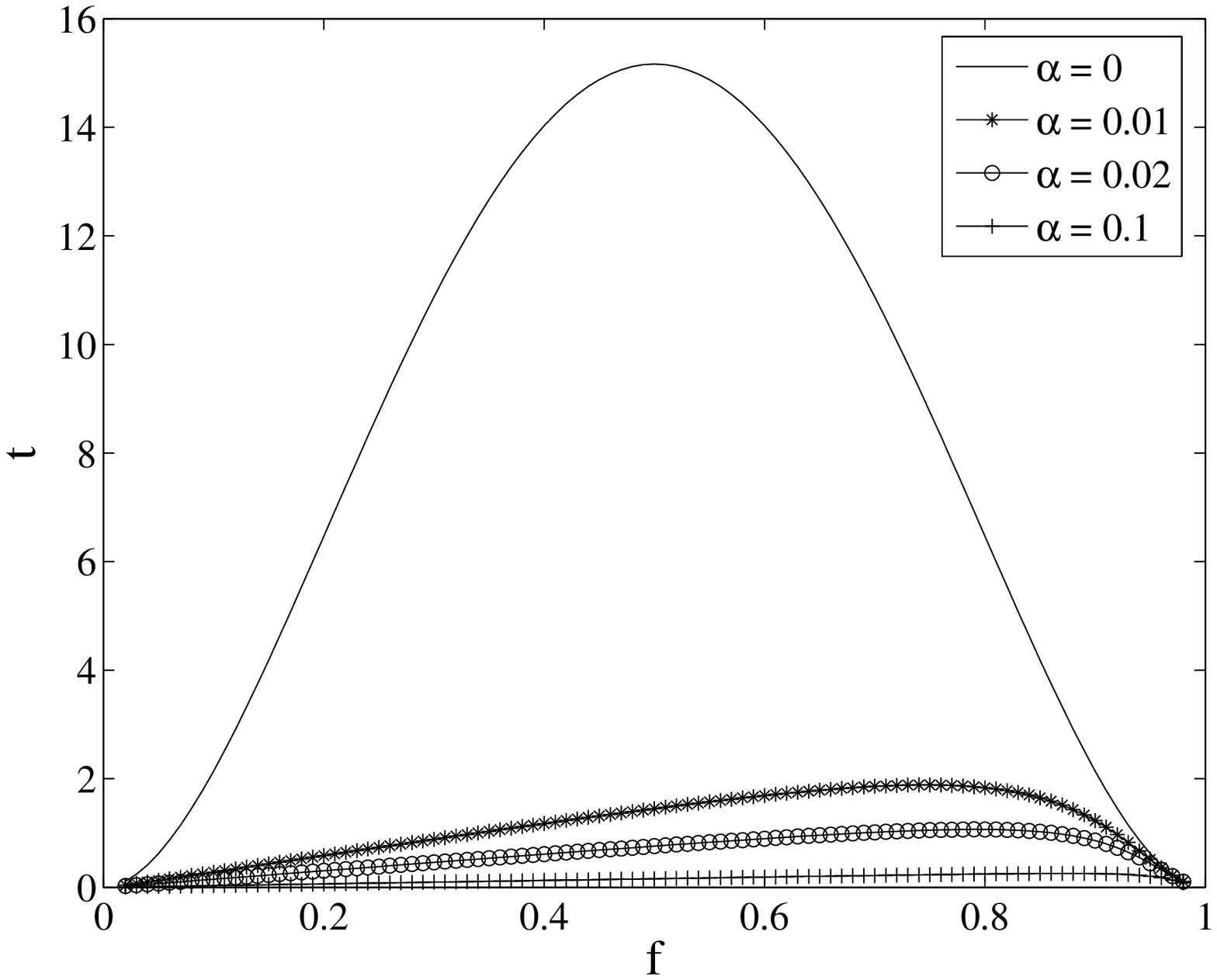}}
   %\label{fig:stability_n10000}
  \end{minipage}
\caption{} \label{stability_n10000}
%\caption{ Effect of degree of ionization ($\alpha$) on the stability limit in polyelectrolytic diblock
%melt: plots correspond to $N = 10,000$ and $\alpha = 0, 0.01, 0.05, 0.1 $.}
 \end{center}
\end{figure}

Having written structure factors in terms of $t$, the stability limit is calculated using 
\begin{eqnarray}
\frac{\delta S^{-1}(k)}{\delta k}\mid _{k=k^{\star},{t=t^{\star}}} &=& 0 \label{eq:stab1}\\
S^{-1}(k)\mid_{k=k^{\star},{t=t^{\star}}} &=& 0 
\end{eqnarray}
where $S^{-1}(k)$ is given in Eqs. (~\ref{eq:struct} - ~\ref{eq:chargestruct}). Solving these equations for $t^{\star}$
\begin{equation}
t^{\star} = \frac{-L+ \sqrt{L^{2}-4 P R}}{2 P} >0,
\label{eq:stability}
\end{equation}
where
\begin{eqnarray}
\quad P &=& 60 \pi x^{\star}A(x^{\star})\\
L &=& 10 \pi f N Z_{c}^{2}A(x^{\star})- 6 x^{\star}+ 10 \pi Z_{A}^{2}\alpha N\\
R &=& -f N Z_{c}^{2} \\
A(x^{\star})&=& Q(x^{\star})
\end{eqnarray}
In writing these equations, we have taken $l=1$. Function $Q$ appearing in these equations has already been defined in Eq. (~\ref{eq:q_func}). Using Eq. (~\ref{eq:stab1}), the wavevector at the stability limit is given by solving the equation
\begin{equation}
\frac{\delta Q(x)}{\delta x}\mid _{x= x^{\star}}= \frac{6 Z_{A}^{2}\alpha N t^{\star}}{(6 x^{\star} t^{\star} + f N Z_{c}^{2})^{2} }. \label{eq:dstable}
\end{equation}
First, Eq. (~\ref{eq:dstable}) is solved for $x^{\star}$ and then $t^{\star}$ is calculated using Eq. (~\ref{eq:stability}). Effect of $\alpha$ and $N$ on the stability limit is shown in Figs. ~\ref{stability_n1000} - ~\ref{stability_n10000}. 
$x^{\star}$ obtained by using Eq. (~\ref{eq:dstable}) for different values of $N$ is shown in Fig. ~\ref{xvsn}.      
%The period ($D = 2\pi/k^{\star}$) obtained by using Eq. (~\ref{eq:dstable}) for different values of $N$ is shown in Fig. ~\ref{periodn}.      

\subsubsection{Ordered Structures}

To derive free energy expressions for different ordered structures, we employ the method used by Leibler\cite{leibler80}. Following Leibler, we expand the free energy expression in terms of the order parameter up to fourth order. Taking advantage of the fact that in WSL (near stability limit), important fluctuations in polymer densities are those with the wavevector $k=k^{\star}$, we approximate the order parameter by a sum of plane waves, each having the wavevector $k=k^{\star}$. 
Using this expression for the order parameter, free energy density becomes
\begin{eqnarray}
\delta F_{n} &=& \frac{N(F-F_{0})}{\Omega k_{B}T} = 2 N l^{3}(\chi_{s}-\chi)\phi_{n}^{2}-\zeta_{n}\phi_{n}^{3}+ \eta_{n}\phi_{n}^{4} \label{eq:free_order1}\\
2 \chi_{s}l^{3}&=& Q(x^{\star})+ \frac{Z_{A}^{2}\alpha N l^{3}}{6 x^{\star} t + f N Z_{c}^{2} }\quad \mbox{and} \quad \chi= \chi_{AB}
\end{eqnarray}
where the value of $n$ corresponds to the morphology being studied. To be specific, $n=1,3$ and $6$ correspond to lamellar, hexagonally close packed (HCP) cylinder and body centred cubic (BCC) spherical morphology, respectively. Functions $\Gamma_{3}, \Gamma_{4}$ and coefficients $\zeta_{n}, \eta_{n}$ were calculated by Leibler\cite{leibler80,typoleib80} ( summarized in Table ~\ref{tab:t1} ). Further, we have adopted the notation used in Ref.\cite{leibler80} for the arguments of the function $\Gamma_{4}$. The coefficients have the property that $\eta_{1}< \eta_{3}<\eta_{6}$ for all $f$  and specifically, for $f=1/2$, $\zeta_{n}=0$. For all the calculations presented in this paper, these coefficients are evaluated at $k = k^{\star}$.
 				     
\subsubsection{Transition Boundaries}
 
By following the Leibler's procedure\cite{leibler80}, the disorder-order transition (DOT) and the order-order transitions (OOT) are studied. Minimizing the free energy density (Eq. (~\ref{eq:free_order1})) with respect to the order parameter, equilibrium order parameter and free energy densities are obtained. Results of these minimizations are presented in Table ~\ref{tab:t2} where $\gamma_{n}$ is given by
\begin{equation}
\gamma_{n}= \left [1-\frac{64 \eta_{n}}{9 \zeta_{n}^{2}}(\chi_{s}-\chi)N \right ]^{1/2}
\end{equation}
 
In order to determine the morphology that evolves at DOT, the free energy density is equated to zero so that $\chi_{n} N$ at DOT is found to be 
\begin{eqnarray}
\chi_{n}N &=& \chi_{s}N- \frac{\zeta_{n}^{2}}{8 \eta_{n}} \label{eq:dot}
\end{eqnarray}
Using the coefficients $\zeta_{n}$ and $\eta_{n}$, it can be shown that BCC $(n=6)$ gives the lowest value for $\chi_{n}N $ (or highest value of $t$). So, the morphology that appears first is BCC. Writing Eq. (~\ref{eq:dot}) for DOT in terms of the reduced temperature $t$, the DOT boundary is given by Eq. (~\ref{eq:stability}) where $A(x^{\star})$ is now given by 
\begin{eqnarray}
A(x^{\star})&=& Q(x^{\star})- \frac{\zeta_{6}^{2}}{4 N \eta_{6}}
\end{eqnarray}
Subscript 6 implies that the morphology is sphere ($n=6$). Similarly, order-order transition boundaries are calculated by equating free energy densities for different morphologies. In general, all the transition boundaries (stability limit, DOT and order-order transitions) are calculated using Eq. (~\ref{eq:stability}), where only $A(x^{\star})$ varies. Mathematical conditions and values of $A(x^{\star})$ for different transition boundaries are summarized in Table ~\ref{tab:t3}.
Function $y$ appearing in Table ~\ref{tab:t3} is the solution of Eq. (V-35) in Leibler's work \cite{leibler80}. Solving these sets of equations, the morphology diagram can be constructed as discussed in Sec. ~\ref{sec:results}.

\subsection{Self-Consistent Field Theory (SCFT)}
\setcounter {equation} {0} \label{sec:SCFT}
Although RPA gives a valuable insight into the physics of the problem, it is only a linear response treatment.
Strictly, this treatment is valid close to the stability limit of homogeneous phase but far from the limit, RPA calculations are not quantitatively correct\cite{dft91,dft93,matsen94}. To go far away from the stability limit, SCFT has been used extensively in the literature\cite{fred04,matsen94,fredbook}. Using standard methods\cite{fredbook}, self-consistent equations are obtained under the saddle point approximation (see Appendix B). 

This numerical technique leads to coupling of full non-linear Poisson-Boltzmann equation with standard modified diffusion equation for the polymer chains. We have solved these sets of equations using an efficient spectral technique\cite{matsen94}. While solving these equations, experimentally found inverse dependence of Flory's $\chi$ parameter has been exploited by using reduced temperature $t$ (Eq. (~\ref{eq:reduced})). We have studied the effect of $\alpha$ on $\chi_{AB}^{\star} N$ and compared with the corresponding RPA calculations. Also, the effect of degree of segregation on the period of lamellar morphology is studied (Fig. ~\ref{all_period}). Monomer densities, counterion densities and electrostatic potential obtained from SCFT calculations are shown in Figs. ~\ref{densities_alp01} - ~\ref{potential_alp01}, respectively..

\section{RESULTS }
\setcounter {equation} {0} \label{sec:results} 

In the previous section , we have provided the necessary equations to describe the microphase separation. Here, we 
present results for salt-free charged-neutral diblock melts, by solving the above equations.
 
\subsection{Stability limit - RPA results}

In Figs. ~\ref{stability_n1000} and ~\ref{stability_n10000}, we have drawn the stability limits for charge/neutral block copolymer salt-free melt 
at different degrees of ionizations for $N = 1000$ and $N = 10,000$, respectively. 
The critical value of the reduced temperature required to induce
microphase separation decreases with an increase in the degree of
ionization. This is in qualitative agreement with limiting laws presented in Sec. ~\ref{sec:RPA} 
and the already established
concept that the effective Flory's $\chi$ parameter
decreases with an increase in degree of ionization for
polyelectrolytes\cite{muthu021}. Unlike the neutral copolymers,
$\chi$ and $N$ for polyelectrolytic diblock copolymers are independent
parameters that govern the phase behaviour. Also, the stability limit
depends on fraction of charged block ($f$) in an unsymmetric fashion. It is to be noted that 
these results are in agreement with the results reported by Marko and Rabin\cite{marko91}. In Ref.\cite{marko91}, temperature dependence of $\chi$ parameter was not taken into account and critical parameters were calculated by choosing a fixed 
value of $l_{B}/l$. 
The method of calculations used by Marko and Rabin was similar to the one presented in 
Appendix A.

\subsection{Period of lamellar phase ($f = 1/2$)}

\begin{figure}[h]
    \begin{center}
    \vspace*{1.0cm}
      \begin{minipage}[c]{15cm}
        \includegraphics[width=15cm]{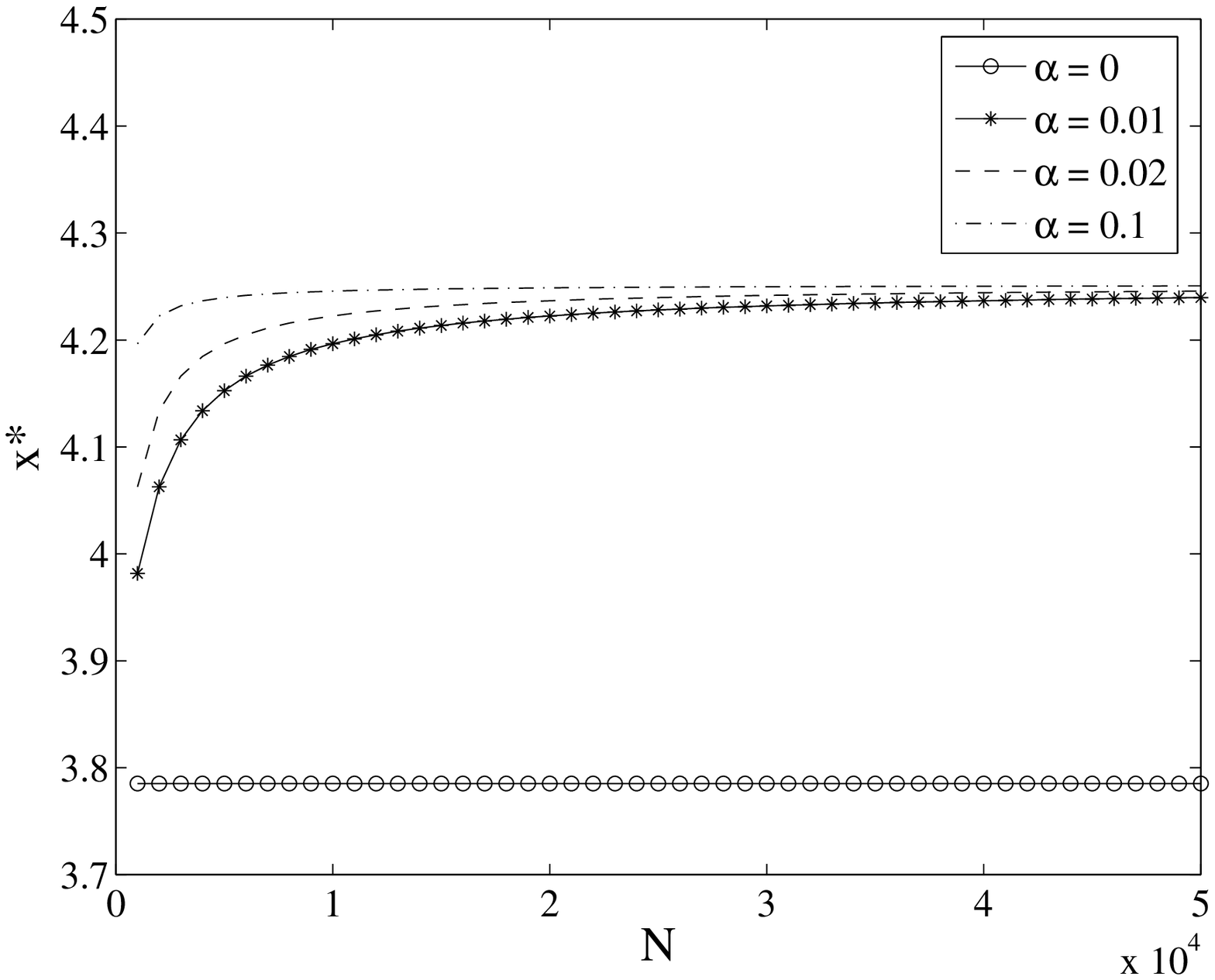}
  %  \label{fig:compare_stab}
    \end{minipage}
\caption{} \label{xvsn}
%\caption{ Comparison of calculated stability limit with Ref\cite{marko91}: $N=10,000\, \mbox{and}\, \alpha=0.05$.}
\end{center}
\end{figure}

In RPA, near the stability limit, the period of an ordered structure is approximated by $D = 2 \pi/k \simeq  2 \pi / k^{\star}$.  It is well known that the mean field theory\cite{leibler80} for neutral block copolymer predicts $\chi_{AB}^{\star}N = 10.495$ and $x^{\star} = 3.7852$ at $f = 1/2$. 
Also, the period shows $1/2$ power law dependence on $N$ in WSL (i.e. $x = \, \mbox{constant}$) 
as long as the wavevector dependence of higher order terms in free energy expression is 
suppressed\cite{dft91,dft93,matsen94}. Physically, this means that block copolymer chains are 
obeying the Gaussian statistics for chain conformations.
Experimentally, there are deviations from this power law because of chain stretching\cite{bates90}.  
As shown in Fig. ~\ref{xvsn}, same power law dependence is obtained using RPA for
polyelectrolytic block copolymers when $N$ is large, but the period ($D \sim \sqrt{N/x^{\star}}$)
for a given $N$ is smaller than that for an equivalent neutral copolymer system.
This effect has been seen by other researchers also\cite{marko91}.
The decrease in period with increase in degree of ionization, is explained in Ref.\cite{marko91} by an argument that counterions need to be rearranged on microphase segregation
and entropy loss is lower if the length scale of fluctuations for these counterions is smaller. 
It is to be stressed that lowering of $D$ with $\alpha$ is {\it not} purely entropic effect. This effect
is an outcome of electrostatic screening due to counterions and hence, includes both energetic as well as entropic contributions. 

\begin{figure}[h]
   \begin{center}
    \vspace*{1.0cm}
      \begin{minipage}[c]{15cm}
    \includegraphics[width=15cm]{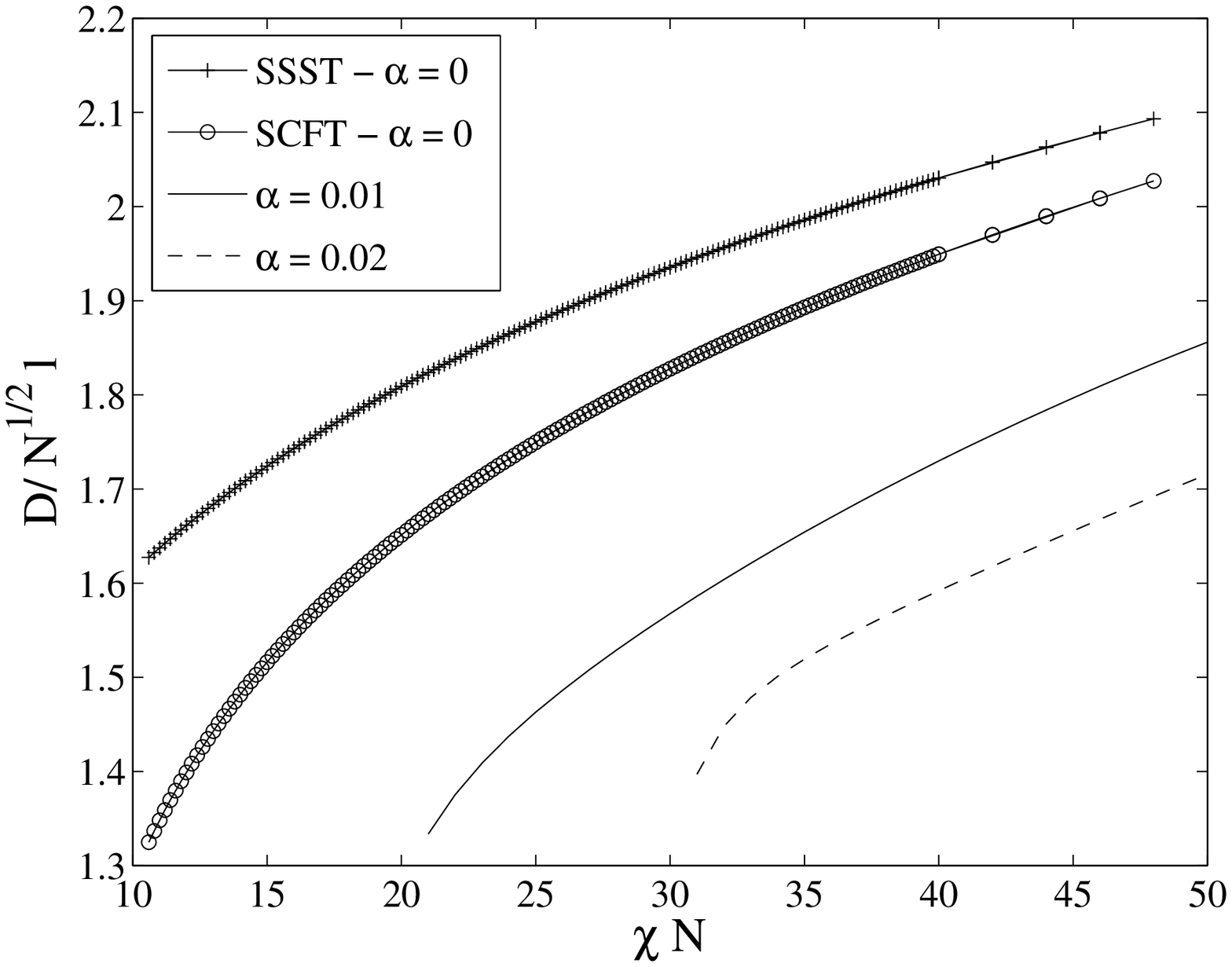}
     \end{minipage}
\caption{} \label{all_period}
%\caption{SCFT Calculations - Effect of degree of segregation on period of lamellae  ($f = 1/2, N = 1000$). Semenov's Strong Segregation Theory (SSST)\cite{semenov84}$^{,}$\cite{matsen94} which predicts $D/\left(N^{1/2} l\right ) = 2 \left (8 \chi N /3 \pi^{4} \right )^{1/6}$ is also drawn for comparison purposes. }
\end{center}
\end{figure}

At present, we are not aware of any experimental data on the period of charged-neutral block copolymer. 
Nevertheless we expect the polyelectrolyte chains to be non-Gaussian and $D$ to deviate strongly from $N^{1/2}$
power law in the case of charged systems.  
SCFT has been quite successful in predicting the period of ordered structures for neutral
copolymers. Expecting that SCFT results are valid for weakly charged polyelectrolyte copolymers, results 
obtained from SCFT calculations are plotted in Fig. ~\ref{all_period} for lamellar phase ($f = 1/2, N = 1000$).
Lowermost point in the plots of Fig. ~\ref{all_period} corresponds to $\chi_{AB}^{\star}N$.
By comparing Fig. ~\ref{all_period} with Figs. ~\ref{stability_n1000} and ~\ref{xvsn}, it is clear 
that the RPA calculations for $\chi_{AB}^{\star} N$
and $x^{\star}$, are in good agreement with the corresponding SCFT calculations. 
Analogous to neutral copolymers, it is 
found that $N^{1/2}$ power law is not valid for ordered microstructures.
In addition the qualitative feature that the period of ordered structures is 
lower than its neutral analog, is clearly seen in these plots.  

\subsection{Counterion distribution}
\begin{figure}[h]
   \begin{center}
    \vspace*{1.0cm}
      \begin{minipage}[c]{15cm}
    \includegraphics[width=15cm]{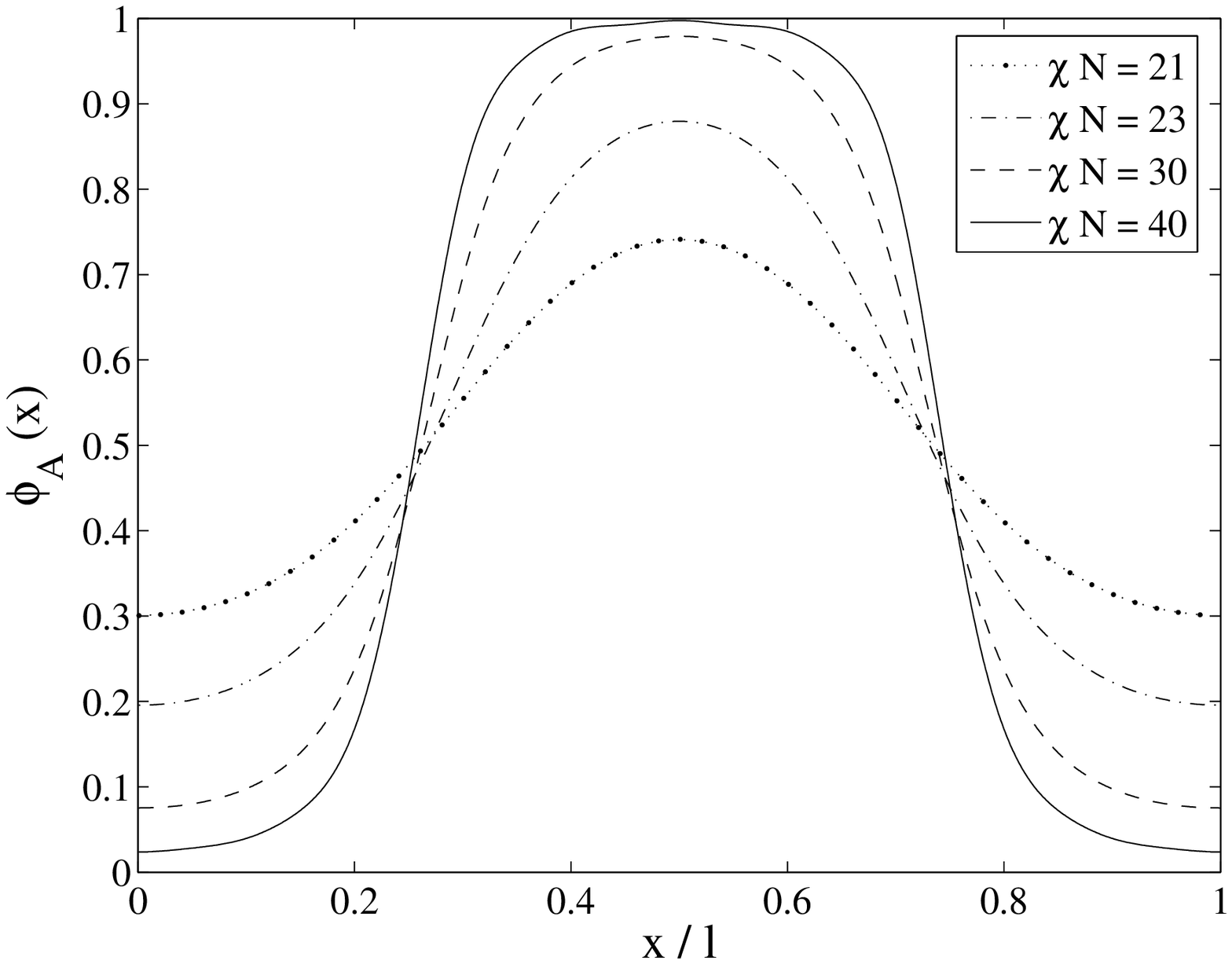}
    \end{minipage}
\caption{}  \label{densities_alp01}
%\caption{Polyelectrolytic block copolymer lamellae ($f = 1/2, \alpha = 0.05, N = 1000$) - monomer densities. }
\end{center}
\end{figure}

\begin{figure}[h]
   \begin{center}
    \vspace*{1.0cm}
      \begin{minipage}[c]{15cm}
    \includegraphics[width=15cm]{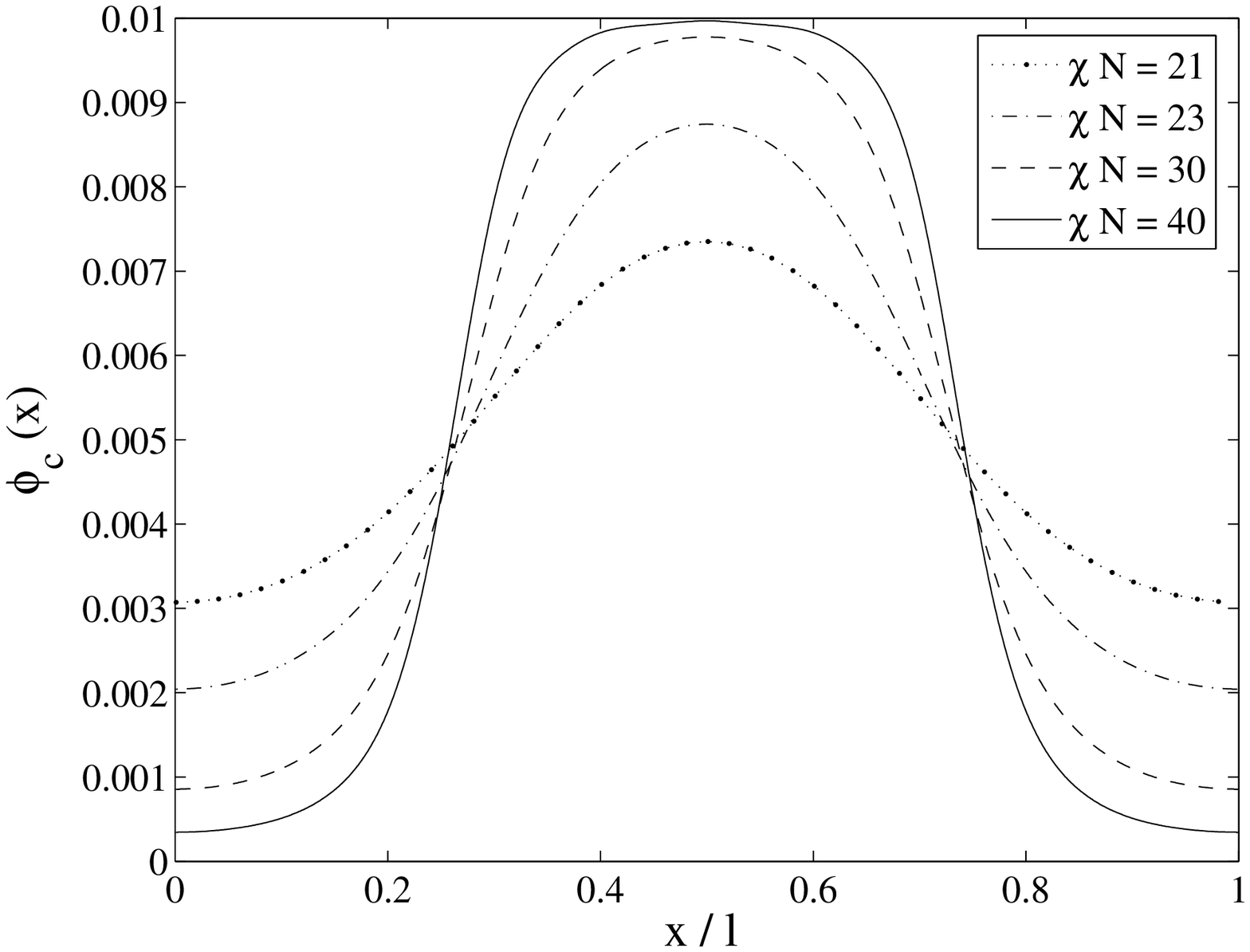}
     \end{minipage}
\caption{} \label{counterion_alp01}
%\caption{Counterion distribution in  lamellar phase ($f = 1/2, \alpha = 0.05, N = 1000$). }
\end{center}
\end{figure}

\begin{figure}[h]
   \begin{center}
    \vspace*{1.0cm}
      \begin{minipage}[c]{15cm}
    \includegraphics[width=15cm]{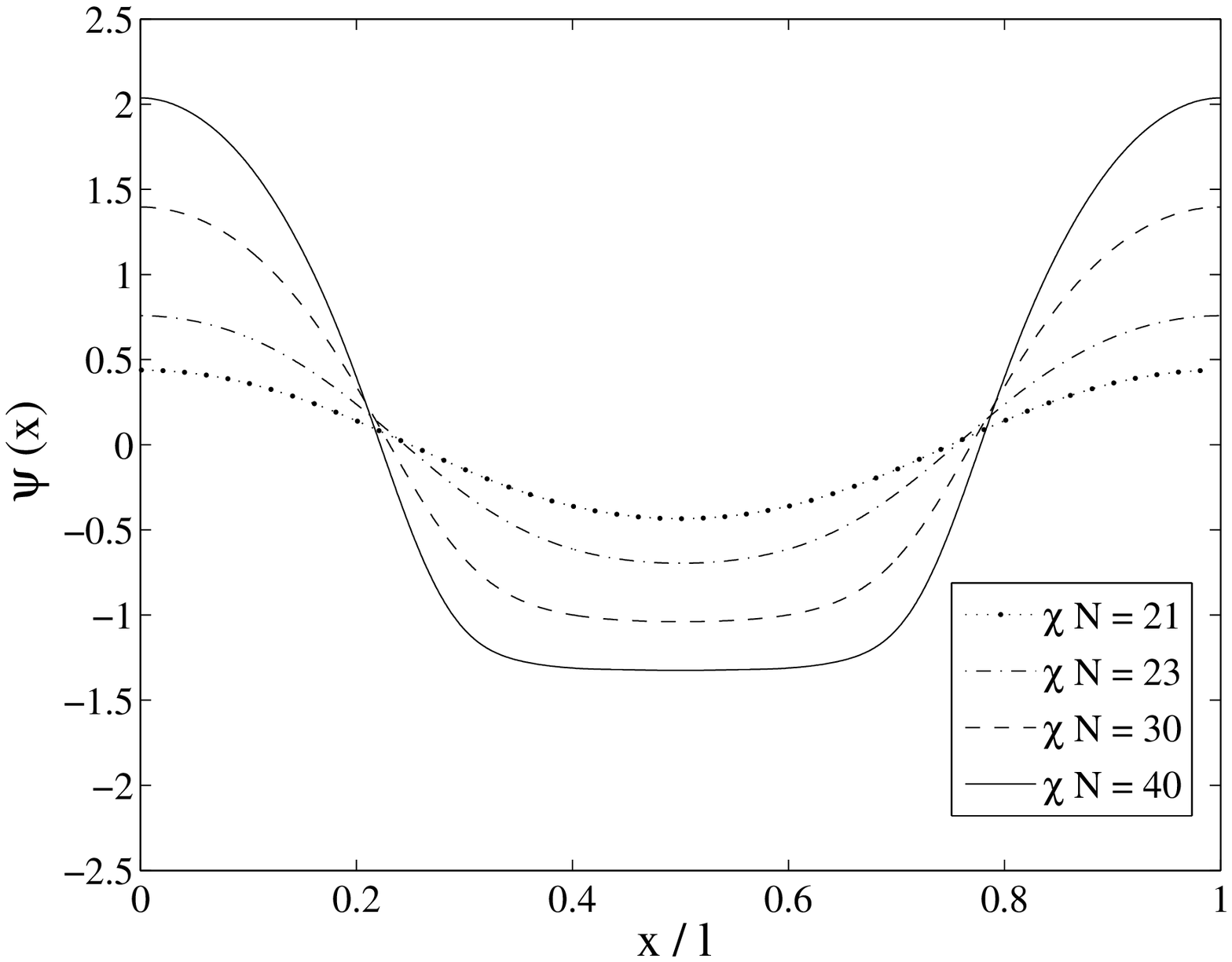}
     \end{minipage} 
\caption{} \label{potential_alp01}
%\caption{Electrostatic potential in  lamellar phase ($f = 1/2, \alpha = 0.05, N = 1000$). }
\end{center}
\end{figure}

\begin{figure}[h]
   \begin{center}
    \vspace*{1.0cm}
      \begin{minipage}[c]{15cm}
    \includegraphics[width=15cm]{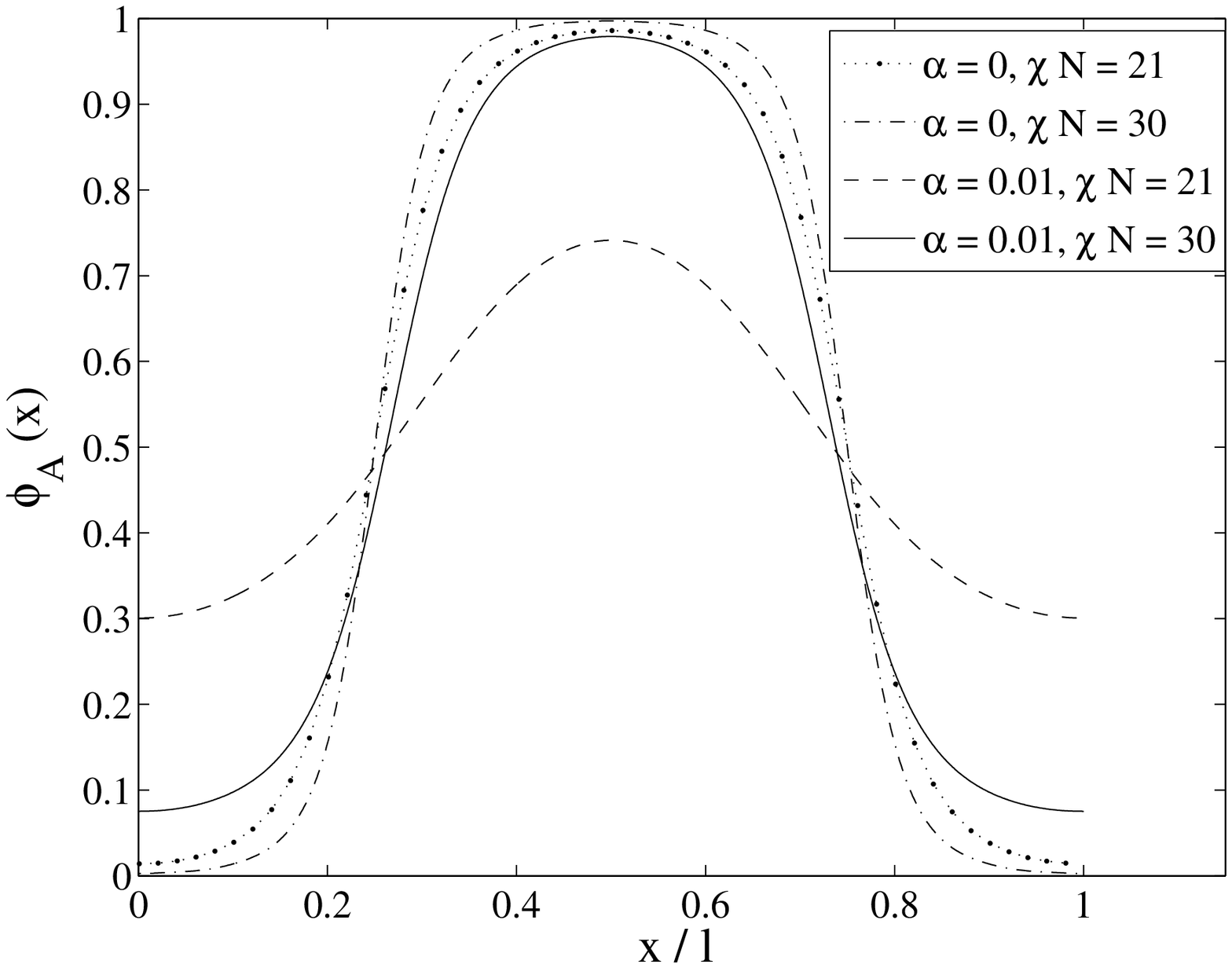}
    \end{minipage}
\caption{} \label{densities_compare}
%\caption{Reduction of effective chemical mismatch ($f = 1/2, \alpha = 0.05, N = 1000$) - comparison between monomer densities. }
\end{center}
\end{figure}

The RPA calculations do not provide counterion distributions. On the other hand, SCFT allows calculations 
of counterion
densities and potential self-consistently. Figs. ~\ref{densities_alp01} - ~\ref{potential_alp01} show monomer 
densities for charged block ($A$),
counterion densities, and the electrostatic potential, respectively. 
The onset of microphase separation 
leads to creation of monomer and counterion density waves (Figs. ~\ref{densities_alp01} and ~\ref{counterion_alp01}) because 
of the incompatibility between the blocks and coupling between charged monomer ($A$) and counterions, respectively. From these plots, it can be inferred that in the strong segregation limit ($\chi N \rightarrow \infty$), all the counterions are confined to the charged domains. One of the effects of these 
density waves in the lamellar phases, is the presence of a potential difference between the charged and neutral domains (Fig. ~\ref{potential_alp01}) whose 
magnitude increases with the degree of segregation. For the weakly charged diblock copolymer system studied here, 
total local charge density is close to zero and it is hard to determine the shape of the charge density wave for the system
due to the possible numerical errors. To verify the observation that the effective degree
of segregation is reduced because of the electrostatic interactions (RPA calculations), we have plotted
monomer densities for neutral and charged block copolymer at the same $\chi N$ (Fig. ~\ref{densities_compare}). These plots 
clearly confirm that effective degree of segregation is lower for charged copolymer melt and counterions have a tendency 
to drive the system towards homogeneous phase.

\subsection{Morphology diagram for charged-neutral diblock copolymer}
\begin{figure}[h]
  \begin{center}
      \vspace*{1.0cm}
      \begin{minipage}[c]{15cm}
     \includegraphics[width=15cm]{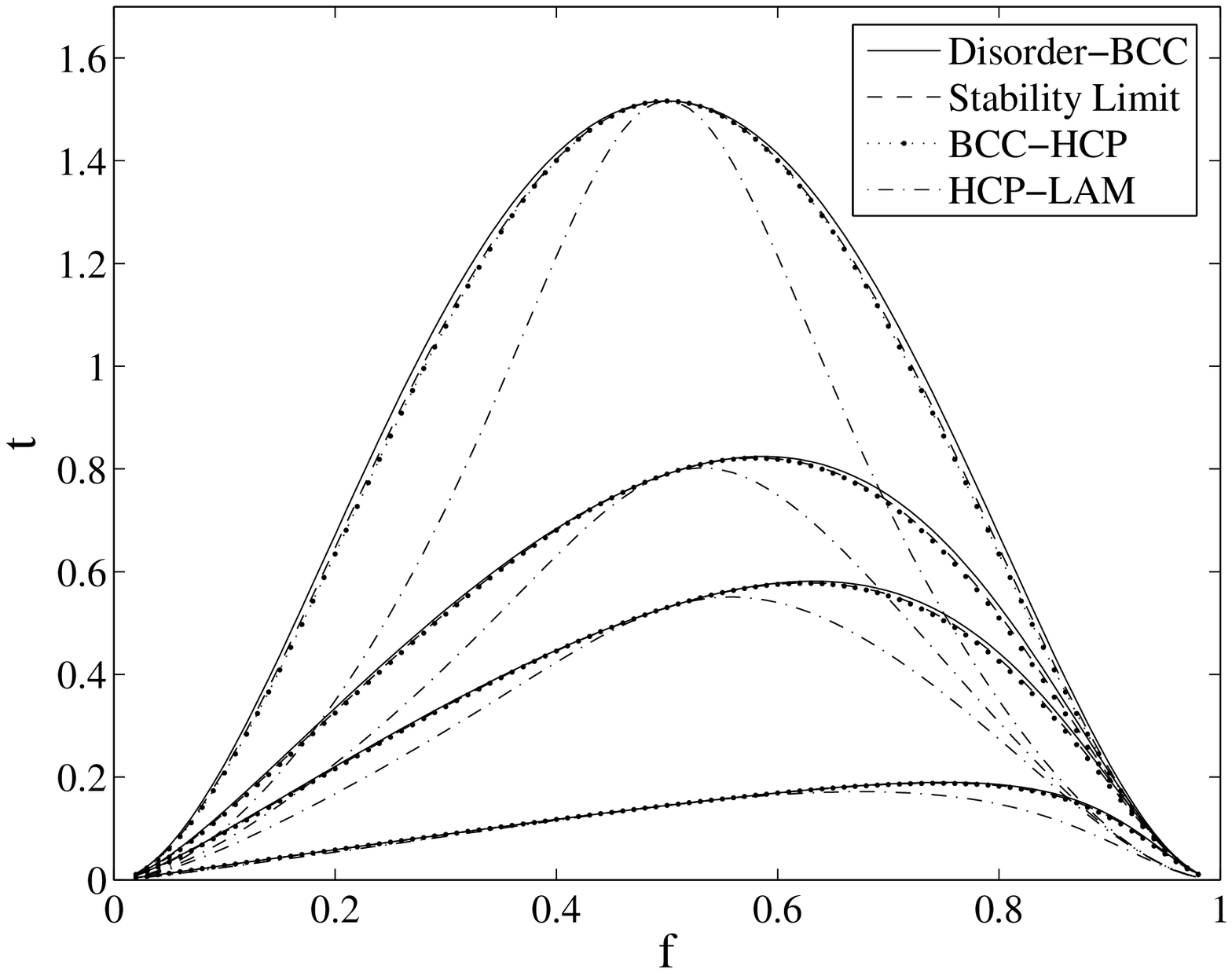}
     \end{minipage}
\caption{} \label{morphology_n1000}
% \caption{ RPA Calculations - Morphology diagram for polyelectrolytic diblock copolymer: $N = 1000$ and $ \alpha = 0$ for the topmost four boundaries, $ \alpha = 0.01$ for the middle four, $ \alpha = 0.05 $ for the next set and $ \alpha = 0.1 $ for the lowermost four boundaries.}
\end{center}
\end{figure}

\begin{figure}[h]
  \begin{center}
      \vspace*{1.0cm}
      \begin{minipage}[c]{15cm}
     \includegraphics[width=15cm]{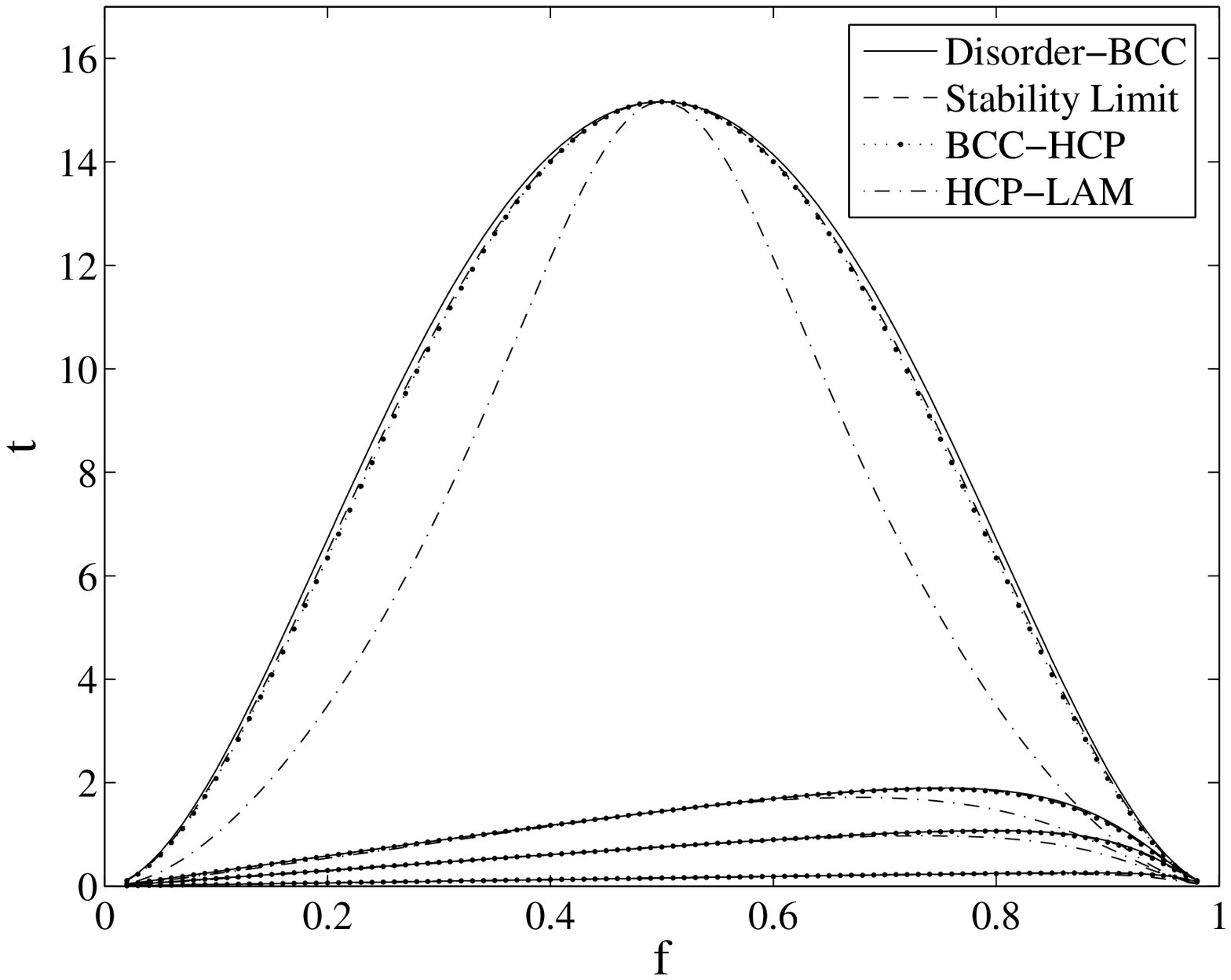}
    \end{minipage}
\caption{} \label{morphology_n10000}
% \caption{ RPA Calculations - Morphology diagram for polyelectrolytic diblock copolymer: $N = 10,000$ and $ \alpha = 0$ for the topmost four boundaries, $ \alpha = 0.01$ for the middle four, $ \alpha = 0.05 $ for the next set and $ \alpha = 0.1 $ for the lowermost four boundaries .}
\end{center}
\end{figure}

The calculation of morphology diagrams using SCFT is a 
computationally intensive task because of the vast parameter space for polyelectrolytic systems.
To get an idea about transition boundaries, we have used RPA calculations presented in 
section ~\ref{sec:RPA}, assuming that only the classical morphologies\cite{leibler80} 
compete in charged-neutral diblock copolymer systems.
Figs. ~\ref{morphology_n1000} and ~\ref{morphology_n10000} show the calculated morphology 
diagrams for different $\alpha$ and $N$. 
We observe that DOT and OOT boundaries are strongly dependent
on $\alpha$ and $N$. The temperature of occurrence of DOT decreases
with an increase in $\alpha$
and increases with an increase in $N$ (analogous to the shift of stability limit - Figs. ~\ref{stability_n1000}, 
~\ref{stability_n10000}) .
Furthermore, these transition boundaries for DOT and OOT are highly asymmetric with respect to $f$. 

\section{CONCLUSIONS} \label{sec:discuss}
We have addressed the microphase separation in charged-neutral diblock copolymer melts in the
 weak segregation limit, by using the RPA and SCFT methods. We have shown in Sections ~\ref{sec:RPA} 
and ~\ref{sec:results} that the critical value of the $\chi$ parameter for microphase separation 
is higher for charged copolymers and of concentration modulation is smaller in comparison with 
neutral copolymers. From 
morphology diagrams, it can be seen that the parameter space for ordered microstructures is reduced 
when degree of ionization of charged block is increased. In other words, charging a block stabilizes the homogeneous phase.

The SCFT results show that the counterions partition themselves 
preferentially within the charged domains. This leads to creation of a potential difference 
between charged and neutral domains.
This process of partitioning is unfavorable both entropically and energetically. Hence, 
the length scale of these partitioning is lower when there are more number of counterions to be partitioned
for the same number of monomers. 

Finally, we summarize the assumptions in obtaining the above results. We have
taken the counterions to be point charges. 
Our treatment can be readily extended to counterions with finite size
by modifying the incompressibility condition and incorporating
excluded volume interactions. The Kuhn segment
lengths for the neutral and charged blocks are taken to be same. It has
been shown that conformational
asymmetry has an effect on the order-order transition boundaries for neutral
copolymers\cite{vava92}. Analogously, there will be an effect on our system as well. 
In the case of polyelectrolytes, electrostatic 
interactions cause stiffening of the chain so that the 
effective segment length\cite{muthu87} depends on various factors such as
$\kappa$, $\alpha$ etc. in a complicated manner. By assuming that the segment length for charged and
neutral blocks to be the same, we implicitly assume that the charged block copolymer under consideration is conformationally 
symmetric and is weakly charged, so that the difference between the effective and bare segment lengths is negligible .
Another important assumption in the present theory is that the position/concentration dependence 
of the dielectric constant $\epsilon$ is suppressed . Recently, the effect of dielectric constants of individual components in a multi-component 
polyelectrolyte\cite{fred04} solution has been presented. In this work, dielectric constant of a component was taken to be 
linearly dependent on the concentration of the component. In principle, dielectric constant 
depends on the concentration of ions in a complex manner\cite{dielectric}.
At present, there is no satisfactory well-established model for the dependence of microscopic dielectric constant on macroscopic density. 
So, we model dielectric constant ($\epsilon$) appearing in the expression for the Bjerrum length as the effective dielectric constant for the mixture of $A, B$ monomers and counterions. Position dependence of dielectric constant will definitely play an important role in the strong segregation limit. However, for melts in WSL, the average value of dielectric constant can be taken as a constant. Further, we have considered only the lamellar, cylindrical and spherical morphologies as the competing structures. Extensions of the present theory for other morphologies are in progress. 

At present we are unaware of any systematic experimental study on charged-neutral copolymer. We hope that the present theoretical work will instigate experimental work on charged block copolymers in the concentrated regime.

 %\vskip0.4cm
%\begin{center}
 %Table 1.
 %Transition Boundaries between Different Morphologies, $l_{B}= 1.1$, $N= 1000$

%\begin{tabular*}{0.75\textwidth}{@{\extracolsep{\fill}} | c | c | c | }
  %\hline
  %$\alpha$ & $\phi_{C}^{S}$ & $\phi_{L}^{C}$  \\
  %\hline
  %Neutral  & 0.215  & 0.355   \\
  %\hline
  %0.01 & 0.18988  & 0.35332    \\
  %\hline
  %0.02  & 0.18869  & 0.35265    \\
  %\hline
  %0.03  & 0.18821 & 0.35251  \\
  %\hline
  %0.04  & 0.18807 & 0.35252    \\
  %\hline
  %0.05  & 0.18791 & 0.35259    \\
  %\hline
  %0.06  & 0.18789 & 0.35265    \\
  %\hline
  %0.07  & 00 & 0.35270    \\
  %\hline
  %0.08  & 00.187890 & 0.35276    \\
  %\hline
  %0.09  & 0.187890 & 0.35281    \\
 % \hline
%\end{tabular*}
%\end{center}
%\vskip0.4cm

\section*{ACKNOWLEDGEMENT}
\setcounter {equation} {0} \label{acknowledgement}
 We acknowledge
financial support from the National Science Foundation (Grant No. DMR-$0605833$) and
the Materials Research Science and Engineering Centre at the University of Massachusetts, Amherst. We are
also grateful to Dr. Chilun Lee for discussions.

\renewcommand{\theequation}{A-\arabic{equation}}
  % redefine the command that creates the equation no.
  \setcounter{equation}{0}  % reset counter
  \section*{APPENDIX A : Integration over positions of small ions }

Here, we present how to compute the integral over positions of small ions. If microsopic charge density at any point in space is defined as 
\begin{eqnarray}
\hat{\rho}_{e}(r)&=& e\left [Z_{A} \alpha \hat{\rho}_{A}(r) +\sum_{m =c,+,- } Z_{m}\hat{\rho}_{m}(r)  \right ]
\end{eqnarray}
where $\hat{\rho}_{A}(r)$ is given by Equation ~\ref{eq:density} and charge density of small ions is defined as
\begin{eqnarray}
\hat{\rho}_{m}(r) &=& \sum_{i =1 }^{n_{m}}Z_{m}\delta \left (r-r_{i} \right ). \label{eq:smalldensity}
\end{eqnarray}

Using these definitions of densities, Equation ~\ref{eq:parti} can be written in the form 
\begin{eqnarray}
       \mbox{exp}\left(-\frac{F} {k_{B}T}\right )= && \frac {1}{n!n_{c}!\prod _{\gamma}
       n_{\gamma}}\int\prod_{j = 1}^{n}D[R_{j}] \int
       \prod_{i=1}^{n_{c} + \sum_{\gamma}n_{\gamma} } dr_{i} \, \mbox{exp} \left [-\frac {3}{2 l}\sum_{j = 1}^{n}
       \int_{0}^{Nl}
       ds_{j}\left(\frac{\partial R_{j}}
       {\partial s_{j}} \right )^{2} \right . \nonumber \\
       & &\mbox{} \left. - l^{3}\int dr  \chi_{AB} \hat{\rho}_{A}(r)\hat{\rho}_{B}(r) - \frac{1}{2}\int {\it dr }\int {\it dr'}\frac{\hat{\rho}_{e}(r)\hat{\rho}_{e}(r')}{\epsilon k_{B}T |r-r'|}  \right ]\nonumber \\
 && \prod_{r}\delta\left[ \hat{\rho}_{A}(r) + \hat{\rho}_{B}(r) -\rho_{0} \right ].\label{eq:new_parti}
\end{eqnarray}
Here,  $\chi_{AB}$ is Flory's chi parameter, which accounts for chemical mismatch between
blocks and is defined as
\begin{equation}
\quad \chi_{AB}l^{3} = w_{AB}-\frac{w_{AA}+w_{BB}}{2}
\end{equation}

Terms corresponding to excluded volume interactions can be treated in the usual way\cite{fredbook}. Here we present 
the treatment of electrostatic terms. We introduce field variables to go from particles to fields by defining $\hat{\phi}_{p}(r)=\frac{\hat{\rho_{p}}(r)}{\rho_{0}}\quad \mbox{where}\quad p = A,B,c,+,-$ (i.e. $p$ corresponds to all components in system i.e. $A$-block,$B$-block,counterions and coions). Before proceeding, we introduce the identity
\begin{eqnarray}
1 &=& \int D[\phi_{p}(r)]\prod_{r} \delta\{\phi_{p}(r)-\hat{\phi_{p}}(r)\} \\
  &=& \int D[\phi_{p}(r)]\int D[w_{p}(r)]\,\mbox{exp}\,\left [{i\rho_{0}\int dr w_{p}(r)\{\phi_{p}(r)-\hat{\phi}_{p}(r) \}}\right ].
\end{eqnarray}
Here, $i$ is purely imaginary number and to avoid confusion, we are going to use the notation $iw_{p}(r)\rightarrow w_{p}(r)$ for all purely imaginary fields. In this notation, Equation ~\ref{eq:new_parti} can be written in the form  
\begin{eqnarray}
\mbox{exp}\left(-\frac{F} {k_{B}T}\right )& = & \int \prod_{p} D[\phi_{p}(r)] D[w_{p}(r)]D[\eta(r)] \,\mbox{exp}\, \left [- \frac{H}{k_{B}T}\right ] \label{eq:new_parti1}
\end{eqnarray}

\begin{eqnarray}
\mbox{exp}\left(-\frac{H} {k_{B}T}\right )& = & \frac {\tau^{n} \Omega^{n_{c}+ \sum_{\gamma}n_{\gamma}}}{n!n_{c}!\prod _{\gamma}n_{\gamma}}\,\mbox{exp}\,\left [ -  l^{3}\rho_{0}^{2}\int {\it dr } \chi_{AB} \phi_{A}(r)\phi_{B}(r) - \frac{\rho_{0}^{2}}{2}\int {\it dr }\int {\it dr'}\frac{\rho_{e}(r)\rho_{e}(r')}{\epsilon k_{B}T |r-r'|}\right . \nonumber \\
& &\mbox{}\left . + \rho_{0} \int {\it dr} \left \{\sum_{p} w_{p}(r) \phi_{p}(r)  + \eta(r)\left ( \phi_{A}(r) + \phi_{B}(r)- 1 \right ) \right \}\right ]Q_{AB}^{n}Q_{c}^{n_{c}}Q_{+}^{n_{+}}Q_{-}^{n_{-}} \nonumber \\
&& 
\end{eqnarray}
where  $ i \eta(r) \rightarrow \eta(r)$ is the field introduced to enforce the incompressibility condition at all points and
\begin{eqnarray}
\rho_{e}(r)=e\left [Z_{A}f \alpha \phi_{A}(r) +\sum_{m = c,+,- } Z_{m}\phi_{m}(r)  \right ]
\end{eqnarray}

\begin{equation}
Q_{AB} = \frac{1}{\tau}\int{\it D}[R] \mbox{exp} \left [- \frac {3}{2 \it l} \int_{0}^{\it Nl}{\it ds}\left(\frac{\partial R(s)}
       {\partial s} \right )^{2} - \frac{1}{l}\int_{0}^{\it N_{A}l}{\it ds}\,w_{A}(R(s))- \frac{1}{l} \int_{\it N_{A}l}^{\it Nl}{\it ds}\, w_{B}(R(s))\right ]
\end{equation}
\begin{equation}
Q_{i} = \frac{1}{\Omega}\int {\it dr}\, \mbox{exp}\, \left [ - w_{i}(r)\right ]
\end{equation}
Here, subscript $i\, ( = c, + , - )$ corresponds to all kinds of small ions in the system. $\Omega$ is the total volume and $\tau$ is a normalization constant given by 
\begin{equation}
\tau = \int D[R] \,\mbox{exp}\left[-\frac {3 }{2 \it l} \int_{0}^{\it N l}{\it ds}\left(\frac{\partial R(s)} {\partial s} \right )^{2} \right] 
\end{equation}
Integrals over $w_{p}$ can't be calculated exactly. So, in order to proceed further, integrals over $w_{i}(r)$ are approximated by the maximum value of the integrand  (saddle point approximation). Maximization of the integrand with respect to $w_{i}(r)$ gives
\begin{equation}
\phi_{i}(r)= \frac{n_{i}}{\rho_{0}\Omega Q_{i}}\,\mbox{exp}\left [-w_{i}(r) \right] \Rightarrow w_{i}(r) = - \mbox{ln}\left[\frac{\rho_{0}\Omega Q_{i}}{n_{i}}\right]- \mbox{ln}\left[\phi_{i}(r)\right]
\label{eq:saddle}
\end{equation}
Plugging $w_{i}$ back into Eq. (~\ref{eq:new_parti1}), 
\begin{eqnarray}
\mbox{exp}\left(-\frac{F} {k_{B}T}\right )& = & \int \prod_{p} D[\phi_{p}(r)] D[w_{A}(r)]D[w_{B}(r)] D[\eta(r)] \,\mbox{exp}\, \left [- \frac{H^{\star}}{k_{B}T}\right ]
\end{eqnarray}
\begin{eqnarray}
\mbox{exp}\left(-\frac{H^{\star}} {k_{B}T}\right )& = & \frac {\tau^{n}(\rho_{0})^{- n_{c}- \sum_{\gamma}n_{\gamma}}}{n!n_{c}!\prod _{\gamma}n_{\gamma}}\,\mbox{exp}\,\left [ -  l^{3}\rho_{0}^{2}\int {\it dr } \chi_{AB} \phi_{A}(r)\phi_{B}(r) \right . \nonumber \\
&& - \frac{\rho_{0}^{2}}{2}\int {\it dr }\int {\it dr'}\frac{\rho_{e}(r)\rho_{e}(r')}{\epsilon k_{B}T |r-r'|} + \rho_{0} \int {\it dr} w_{A}(r) \phi_{A}(r) \nonumber \\
&& +  \rho_{0} \int {\it dr} w_{B}(r) \phi_{B}(r) - \rho_{0} \int {\it dr} \sum_{i = c, +}^{-} \phi_{i}(r)\,\mbox{ln}\left [\phi_{i}(r)\right ] \nonumber \\
&& \left . + \rho_{0}\int dr \, \eta(r)\{ \phi_{A}(r) + \phi_{B}(r)- 1 \} \right ]Q_{AB}^{n}. \label{eq:freemarko}
\end{eqnarray}
In order to calculate the integral over $\phi_{i}$, we consider fluctuations of charge densities about the disordered phase\cite{marko91}. For weak fluctuations, the entropic term $\phi_{i} \mbox{ln}\phi_{i}$ can be expanded in powers of these fluctuations. Keeping the leading terms (neglecting cubic and higher order terms), integrals over $\phi_{i}$ become Gaussian and can be easily calculated. So, by writing 
\begin{equation} 
\phi_{i}(r) = <\phi_{i}>  +  \delta \phi_{i}(r) \label{eq:density_fluct}
\end{equation}
, and using $\int dr \, \delta \phi_{i}(r) = 0$ in combination with the global electroneutrality in the homogeneous phase, integrals over $\phi_{i}$ (i.e. small ions) become
\begin{eqnarray}
I &=& \int \prod_{m=1}^{3}D[\delta \phi_{m}]\mbox{exp}\left [-\frac{\rho_{0}^{2}}{2}\int \frac{d^{3}k}{(2\pi)^{3}}\left \{ \sum_{m=1}^{3} \sum_{q=1}^{3}\delta \phi_{m}(k) \left (\frac{\delta_{mq}}{\rho_{0}<\phi_{m}>} + V_{k}^{(0)}Z_{m}Z_{q} \right )\delta \phi_{q}(k) \right . \right . \nonumber \\
&&\left . \left . + 2 \sum_{m=1}^{3} V_{k}^{(0)} \alpha Z_{A}Z_{m}\delta \phi_{A}(k)\delta \phi_{m}(k)  \right \}  \right ]\mbox{exp}\left[ -\frac{\rho_{0}^{2}}{2}\int \frac{d^{3}k}{(2\pi)^{3}}V_{k}^{(0)}\alpha^{2}Z_{A}^{2}\delta \phi_{A}^{2}(k) \right ],\nonumber \\
\mbox{} \label{eq:intmarko}
\end{eqnarray}
where $V_{k}^{(0)}= \frac{4\pi l_{B}}{k^{2}} $, $\delta_{mp}$ is the Kronecker delta and the indices 1,2,3 correspond to counterion from polymer, positive (+) and negative (-) salt ions, respectively. By calculating the Gaussian integrals, the value of $I$ is found to be
\begin{eqnarray}
I = \frac{(2\pi)^{\frac{3}{2}}}{\rho_{0}^{3}\sqrt{c_{1}c_{2}c_{3}}}&&\mbox{exp}\left [-\frac{\Omega}{2}\int \frac{d^{3}k}{(2\pi)^{3}}\mbox{ln}\left [ 1 + V_{k}^{(0)}c_{0} \right ]\right ]\nonumber \\
&& \mbox{exp}\left [-\frac{\rho_{0}^{2}}{2} \alpha^{2} Z_{A}^{2}\int \frac{d^{3}k}{(2\pi)^{3}} \frac{V_{k}^{(0)}}{1+ V_{k}^{(0)}c_{0}} \delta \phi_{A}^{2}(k)\right ] 
\end{eqnarray}
where $c_{m} = \frac{\Omega}{n_{m}}$ and $c_{0}$ is given by
\begin{equation}
c_{0}= \frac{1}{\Omega}\sum_{m = 1,2,3} Z_{m}^{2}n_{m} = \frac{\kappa^{2}}{4 \pi l_{B}} \label{eq:c0}
\end{equation}

Now, using the relation
\begin{equation}
\int_{0}^{\infty} x^{2}\left \{\mbox{ln}\left[1 +\frac{\alpha^{2}}{x^{2}} \right ] - \frac{\alpha^{2}}{x^{2}}  \right \} = - \frac{\pi}{3}\alpha^{3},
\end{equation}
we get
\begin{equation}
\frac{\Omega}{2}\int\frac{d^{3}k}{(2\pi)^{3}}\mbox{ln}(1+ c_{0}V_{k}^{(0)}) = -\frac{\Omega}{12 \pi}\left (4\pi l_{B} c_{0} \right )^{3/2} + \frac{\Omega}{2}\int\frac{d^{3}k}{(2\pi)^{3}}\frac{4 \pi l_{B}c_{0}}{k^{2}}
\end{equation}
Here, the last term is divergent. But the divergence is for $k$ being large (ultraviolet divergence) and we are interested on a length scale corresponding to small $k$. So, this term is neglected. Note that the same result can be obtained by employing the central limit theorem\cite{muthu96}. Now, electrostatic terms are decoupled from the rest of the terms and remaining treatment is the same as that done for the corresponding neutral copolymer\cite{ohta86}.  

\renewcommand{\theequation}{B-\arabic{equation}}
  % redefine the command that creates the equation no.
  \setcounter{equation}{0}  % reset counter
  \section*{APPENDIX B : Self-Consistent Field Theory }

Instead of expanding about average densities, the introduction of another field corresponding to charge density\cite{shi99}$^{,}$\cite{fred04} in Eq. (~\ref{eq:new_parti1}), leads to the following self consistent equations for the {\it salt-free melt} after taking $l = 1$
\begin{eqnarray}
\chi_{AB} N \phi_{B}(r) &=& w_{A}(r) + \eta(r) \label{eq:sad1}\\
\chi_{AB} N \phi_{A}(r) &=& w_{B}(r) + \eta(r) \label{eq:sad2} \\
\phi_{A}(r)+\phi_{B}(r) &=& 1 \label{eq:sad3} \\
\phi_{C}(r)&=& -\frac{Z_{A}f \alpha}{Z_{c} Q_{c}}\mbox{exp}\left [- Z_{c}\psi(r) \right] \label{eq:sad4} \\
\phi_{A}(r)&=&  \frac{\Omega \int_{0}^{f } ds \, q(r,s) q^{\star}(r,1-s)}{\int dr \, q(r,1)} \label{eq:sad5} \\
\phi_{B}(r)&=&  \frac{\Omega \int_{f }^{1} ds \, q(r,s) q^{\star}(r,1-s)}{\int dr \, q(r,1)} \label{eq:sad6} \\
\bigtriangledown_{r}^{2}\psi(r) &=& - 4\pi l_{B} \left [ Z_{c}\phi_{C}(r) + Z_{A}\alpha \phi_{A}(r) \right ] \label{eq:sad7}\\
&& \nonumber \\
\frac{\partial q(r,s) }{\partial s} &=& \left \{ \begin{array}{ll}
\left [\frac{N }{6}\bigtriangledown_{r}^{2}- \left \{ Z_{A}\alpha N \psi(r) + w_{A}(r)\right \} \right ]q(r,s) & s \leq f \\
  &   \\
\left [\frac{N }{6}\bigtriangledown_{r}^{2}- w_{B}(r) \right ]q(r,s) & s \geq f
\end{array} \right . \\
&& \nonumber \\
\frac{\partial q^{\star}(r,t) }{\partial t} &=& \left \{ \begin{array}{ll}
\left [\frac{N }{6}\bigtriangledown_{r}^{2}- \left \{ Z_{A}\alpha N \psi(r) + w_{A}(r)\right \} \right ]q^{\star}(r,t) & t \geq (1-f) \\
  & \\
\left [\frac{N }{6}\bigtriangledown_{r}^{2}- w_{B}(r) \right ]q^{\star}(r,t) & t \leq (1-f)
\end{array}  \right .  
\end{eqnarray}
These equations are to be solved with initial conditions $q(r,0) = 1, q^{\star}(r,0) = 1$ and the free energy expression for the salt-free melt (per chain) becomes
\begin{eqnarray}
\frac{F}{n k_{B}T} & = & - \frac{1}{\Omega}\int {\it dr } \chi_{AB} N \phi_{A}(r)\phi_{B}(r) -  \frac{N }{8 \pi l_{B} \Omega} \int dr |\bigtriangledown_{r}\psi(r)|^{2} +  \frac{1}{\Omega}\int dr \, \eta(r) \nonumber \\
& &   - \left \{ \mbox{ln} \left [ \frac{Q_{AB}}{n} \right ] -\frac{Z_{A}f \alpha N}{Z_{c}} \,\mbox{ln}\left [ \frac{Q_{c}}{n_{c}}\right ]  -  \frac{Z_{A}f \alpha N}{Z_{c}} \left ( \mbox{ln}\Omega + 1 \right ) + \left ( \mbox{ln}\tau + 1 \right ) \right \}
\end{eqnarray}
Here, we have used the notation $i e\psi(r) \rightarrow \psi(r)$ for purely imaginary electrostatic potential and $i w_{p}(r) \rightarrow w_{p}(r)$ for all the purely imaginary fields. Expressions for $Q_{c}, Q_{AB}$ and $\tau$ are given in Appendix A.  The free energy of the homogeneous phase (where all the densities and fields are constant) is given by
\begin{eqnarray}
\frac{F}{n k_{B}T} & = & \chi_{AB} N f (1-f) + \left ( -\frac{Z_{A}}{Z_{c}}f \alpha N \right) \left \{ \,\mbox{ln}\left [ - \frac{Z_{A}}{Z_{c}}f \alpha \right ] - 1 \right \}
\end{eqnarray}

\newpage
\section*{REFERENCES}
\pagestyle{empty} \label{REFERENCES}

\newpage

\section*{FIGURE CAPTION}
\pagestyle{empty}
\begin{description}
\item[Fig. 1.:] Effect of degree of ionization ($\alpha$) on the stability limit in polyelectrolytic diblock
melt: plots correspond to $N = 1000$ and $\alpha = 0, 0.01, 0.02, 0.1 $.
\end{description}

\begin{description}
\item[Fig. 2.:] Effect of degree of ionization ($\alpha$) on the stability limit in polyelectrolytic diblock
melt: plots correspond to $N = 10,000$ and $\alpha = 0, 0.01, 0.02, 0.1 $.
\end{description}

%\begin{description}
%\item[Fig. 3.:] Comparison of calculated stability limit with Ref\cite{marko91}: $N=10,000\, \mbox{and}\, \alpha=0.05$.
%\end{description}
\begin{description}
\item[Fig. 3.:] RPA Calculations - Effect of degree of polymerization ($N$) and degree of ionization ($\alpha$) on critical parameter $x^{\star}$ for ($f=\frac{1}{2}$) in WSL : $\alpha = 0, 0.01, 0.02, 0.1 $.
\end{description}

%\begin{description}
%\item[Fig. 3.:] RPA Calculations - Effect of degree of polymerization ($N$) and degree of ionization ($\alpha$) on period $D \simeq 2 \pi /k^{\star}$ for ($f=\frac{1}{2}$) in WSL : $\alpha = 0, 0.01, 0.02, 0.1 $ ($D \sim  N^{1/2}$).
%\end{description}

\begin{description}
\item[Fig. 4.:] SCFT Calculations - Effect of degree of segregation on period of lamellae  ($f = 1/2, N = 1000$). Semenov's Strong Segregation Theory (SSST)\cite{semenov84}$^{,}$\cite{matsen94} which predicts $D/\left(N^{1/2} l\right ) = 2 \left (8 \chi N /3 \pi^{4} \right )^{1/6}$ is also drawn for comparison purposes.
\end{description}

\begin{description}
\item[Fig. 5.:] Polyelectrolytic block copolymer lamellae ($f = 1/2, \alpha = 0.01, N = 1000$) - monomer densities.
\end{description}

\begin{description}
\item[Fig. 6.:] Counterion distribution in  lamellar phase ($f = 1/2, \alpha = 0.01, N = 1000$).
\end{description}

%\begin{description}
%\item[Fig. 8.:] Local charge density in  lamellar phase ($f = 1/2, \alpha = 0.05, N = 1000$)
%\end{description}

\begin{description}
\item[Fig. 7.:] Electrostatic potential in  lamellar phase ($f = 1/2, \alpha = 0.01, N = 1000$).
\end{description}

\begin{description}
\item[Fig. 8.:] Reduction of effective chemical mismatch ($f = 1/2, N = 1000$) - comparison between monomer densities.
\end{description}

\begin{description}
\item[Fig. 9.:] RPA Calculations - Morphology diagram for polyelectrolytic diblock copolymer: $N = 1000$ and $ \alpha = 0$ for the topmost four boundaries, $ \alpha = 0.01$ for the middle four, $ \alpha = 0.02 $ for the next set and $ \alpha = 0.1 $ for the lowermost four boundaries.
\end{description}

\begin{description}
\item[Fig. 10.:] RPA Calculations - Morphology diagram for polyelectrolytic diblock copolymer: $N = 10,000$ and $ \alpha = 0$ for the topmost four boundaries, $ \alpha = 0.01$ for the middle four, $ \alpha = 0.02 $ for the next set and $ \alpha = 0.1 $ for the lowermost four boundaries. 
\end{description}

\clearpage
\vskip0.4cm
\begin{table}
\begin{center}
\begin{tabular*}{0.75\textwidth}{@{\extracolsep{\fill}} | c | c | c | }
  \hline
  Morphology & $\zeta_{n}$  & $\eta_{n}$  \\
  \hline
  Lamellar & $\zeta_{1}= 0$  & $\eta_{1} = \frac{N}{4} \Gamma_{4}(0,0) \quad $    \\
  \hline
  Cylinder &  $\zeta_{3}= -(2/3 \sqrt{3})N \Gamma_{3}$ & $\eta_{3} = \frac{N}{12}( \Gamma_{4}(0,0)+ 4 \Gamma_{4}(0,1))$ \\
  \hline
  Sphere &  $\zeta_{6} = -(4/3 \sqrt{6})N \Gamma_{3}$  & $\eta_{6} = \frac{N}{24}( \Gamma_{4}(0,0)+ 8 \Gamma_{4}(0,1)$ \\
                      $\mbox{}$              &  $ \quad + 2 \Gamma_{4}(0,2)+ 4\Gamma_{4}(1,2)) $  & \mbox{}    \\
  \hline
\end{tabular*}
\caption{\label{tab:t1} Coefficients $\zeta_{n}$ and $\eta_{n}$ calculated by Leibler\cite{leibler80}}
\end{center}
\end{table}
\vskip0.4cm

\clearpage 
%newpage

\vskip0.4cm
\begin{table}
\begin{center}
\begin{tabular*}{0.95\textwidth}{@{\extracolsep{\fill}}|c|c|c|}
  \hline
  Morphology  & Order Parameter $(\bar{\phi_{n}} )$ & Free Energy Density ($\delta F_{n}$)\\
  \hline
  Disorder - Order Transition  & $\zeta_{n}/\left (2 \eta_{n}\right ) $ & 0 \\ \hline
  Lamellar  & $\sqrt{(\chi - \chi_{s})N/\eta_{1}} $  & $- N^{2}(\chi_{s}-\chi)^{2}/\eta_{1} $\\ \hline
  Sphere, Cylinder & $3 \zeta_{n}(1 + \gamma_{n})/\left (8 \eta_{n} \right )$ & $27 \zeta_{n}^{4}(1 + \gamma_{n})^{3}(1-3 \gamma_{n})/\left (4096 \eta_{n}^{3} \right ) $  \\  \hline
\end{tabular*}
\caption{\label{tab:t2} Equilibrium order parameters and free energy densities }
\end{center}
\end{table}
\vskip0.4cm

\clearpage

\vskip0.4cm
\begin{table}
\begin{center}
 \begin{tabular*}{0.95\textwidth}{@{\extracolsep{\fill}}|c|c|c|}
  \hline
  Transition Boundary  & Mathematical Conditions  & $A(x^{\star})$\\
 \hline
Stability Limit & $\frac{\delta S^{-1}(k)}{\delta k}\mid _{k=k^{\star},{t=t^{\star}}} = 0$ & $Q(x^{\star})$\\
& $S^{-1}(k)\mid_{k=k^{\star},{t=t^{\star}}} = 0$ & \\
\hline
  $\mbox{}$  & $  \frac{\partial(\delta F_{n})}{\partial \phi_{n}}\mid _{\phi_{n} = \bar{\phi_{n}}}= 0 $ & $\mbox{}$ \\            Disorder - Order & $ \frac{\partial^{2}(\delta F_{n})}{\partial \phi_{n}^{2}}\mid _{\phi_{n} = \bar{\phi_{n}}}> 0 $ & $Q(x^{\star})- \zeta_{6}^{2}/\left ( 4 N\eta_{6} \right )$ \\
                    $\mbox{}$ &   $ \delta F_{n}(\bar{\phi_{n}},t_{n}) = 0 $ &  $\mbox{}$   \\
  \hline
  Sphere - Cylinder & $\delta F_{6}= \delta F_{3}$  &  $Q(x^{\star})+ 2 y/N$   \\  \hline   Cylinder - Lamellar  & $\delta F_{3} = \delta F_{1}$ & $Q(x^{\star})+ 9\zeta_{3}^{2}(\gamma_{3}^{2} - 1)/\left (32 N\eta_{3}\right )$\\
  \hline
\end{tabular*}
\caption{\label{tab:t3} Description of different transition boundaries}
\end{center}
\end{table}
\vskip0.4cm

\clearpage

\vspace*{1.0cm}

\newpage
\vspace*{1.0cm}

\newpage
\vspace*{1.0cm}

\newpage
%\vspace*{1.0cm}
%\begin{figure}[h]
%\begin{center}  
    % \vspace*{1.0cm}
   %   \begin{minipage}[c]{15cm}
  %   \includegraphics[width=15cm]{periodn.eps}
 %       \end{minipage}
%\caption{} \label{periodn}
%\caption{RPA Calculations - Effect of degree of polymerization ($N$) and degree of ionization ($\alpha$) on period $D \simeq 2 \pi /k^{\star}$ for ($f=\frac{1}{2}$) in WSL : $\alpha = 0, 0.01, 0.02, 0.1 $ ($D \sim  N^{1/2}$)  .}
%\end{center}
%\end{figure}

%\newpage
\vspace*{1.0cm}

\newpage
\vspace*{1.0cm}

\newpage
\vspace*{1.0cm}

%\newpage
%\vspace*{1.0cm}
%\begin{figure}[h]
   %\begin{center}
    %\vspace*{1.0cm}
    %  \begin{minipage}[c]{15cm}
   % \includegraphics[width=15cm]{charge_alp05.eps}
   % \includegraphics[width=15cm]{charge_alp05_corrected.eps}
  %  \label{fig:charge}
 %   \end{minipage}
%\caption{}
%\caption{Local charge density in  lamellar phase ($f = 1/2, \alpha = 0.05, N = 1000$). }
%\end{center}
%\end{figure}

\newpage
\vspace*{1.0cm}

\newpage
\vspace*{1.0cm}

\newpage
\vspace*{1.0cm}

\newpage
\vspace*{1.0cm}

\end{document}